\documentclass{article}
\usepackage{multicol}

\usepackage{arxiv}

\usepackage{cite}
\usepackage{hyperref}       %
\usepackage{url}            %
\usepackage{booktabs}       %
\usepackage{amsfonts}       %
\usepackage{nicefrac}       %
\usepackage{microtype}      %
\usepackage{lipsum}
\usepackage{graphicx}

\usepackage{amsmath}
\usepackage{lmodern}%
\usepackage{textcomp}%
\usepackage{lastpage}%
\usepackage{float}%
\usepackage{xcolor}
\usepackage[export]{adjustbox}
\usepackage{listings}
\usepackage{multirow}
\usepackage{longtable}
\usepackage{authblk}
\usepackage{lscape}
\usepackage{array}
\usepackage{caption}
\usepackage{subcaption}

\begin{document}
\title{Natural Aging and Vacancy Trapping in Al-6xxx}
\author[1]{Abhinav C. P. Jain}
\author[2]{M. Ceriotti}
\author[1]{W. A. Curtin}
\affil[1]{Institute of Mechanical Engineering, École Polytechnique Fédérale de Lausanne, CH-1015, Lausanne, Switzerland}
\affil[2]{Institute of Materials Science, École Polytechnique Fédérale de Lausanne, CH-1015, Lausanne, Switzerland}

\maketitle

\begin{abstract}
Undesirable natural aging (NA) in Al-6xxx delays subsequent artificial aging (AA) but the size, composition, and evolution of clustering are challenging to measure.  Here, atomistic details of early-stage clustering in Al-1\%Mg-0.6\%Si during NA are studied computationally using a chemically-accurate neural-network potential.   Feasible growth paths for the preferred $\beta''$ precipitates identify: dominant clusters differing from $\beta''$ motifs; spontaneous vacancy-interstitial formation creating 14-18 solute atom $\beta''$-like motifs; and lower-energy clusters requiring chemical re-arrangement to form $\beta''$ nuclei.  Quasi-on-lattice kinetic Monte Carlo simulations reveal that 8-14 solute atom clusters form within 1000 s but that growth slows considerably due to vacancy trapping inside clusters, with trapping energies of 0.3-0.5 eV.  These findings rationalize why cluster growth and alloy hardness saturate during NA, confirm the concept of ''vacancy prisons", and suggest why clusters must be dissolved during AA before formation of $\beta''$.  This atomistic understanding of NA may enable design of strategies to mitigate negative effects of NA.
\end{abstract}

\section{Introduction}

The Al-6XXX alloys containing Mg and Si possess a high strength to weight ratio and thus find extensive automotive and structural applications \cite{Miller2000}. The high strength is achieved via a controlled precipitation process (artificial aging (AA)) \cite{Hirsch2016}. The typical aging sequence is (i) solutionizing at $\sim$800 K, (ii) quenching to room temperature to retain a high (non-equilibrium) vacancy concentration, and (iii) annealing at 443 K for $\sim 10^4 - 10^5$ s  to precipitate the preferential $\beta''$ Mg-Si structures that are responsible for hardening at peak aging \cite{Anderson1998,Pogatscher2011}. Longer aging leads to the formation of the thermodynamically-stable $\beta$ phase that leads to a weaker alloy.  

\begin{figure}[htbp]
	\centering
	\subfloat[Hardness curves of AA 6061 alloy \cite{Pogatscher2011}.]{\label{fig:a}\includegraphics[width=0.4\linewidth]{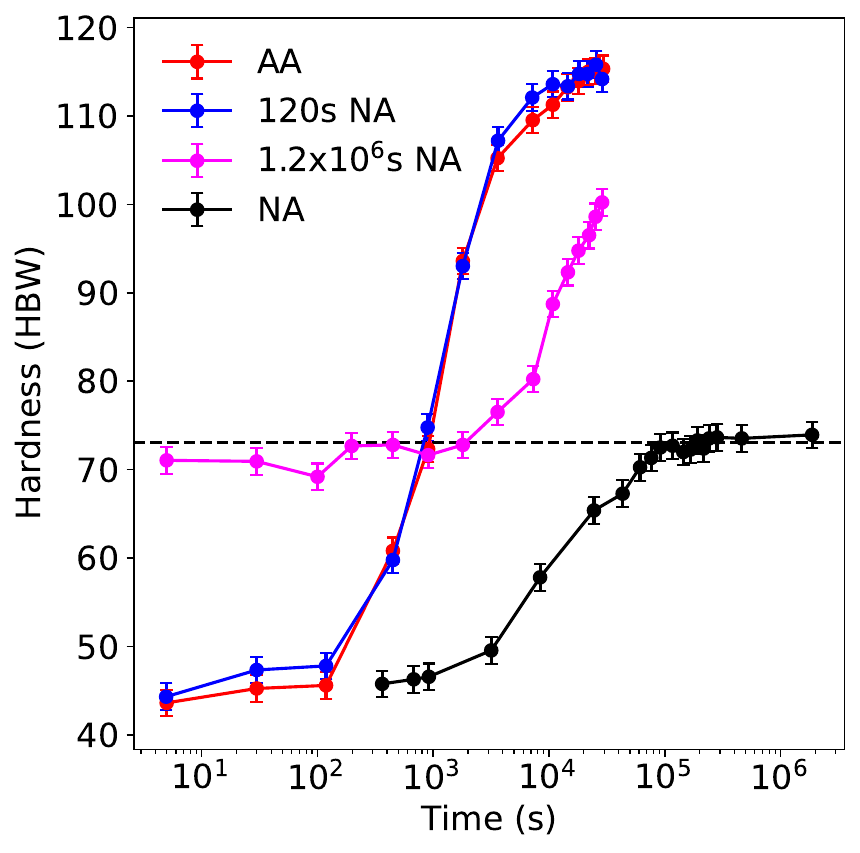}
    }\qquad
	\subfloat[Excess heat flow (DSC heating curves) \cite{Dumitraschkewitz2019Size-dependentAlloys}.]{
        \label{fig:b}\includegraphics[width=0.5\linewidth]{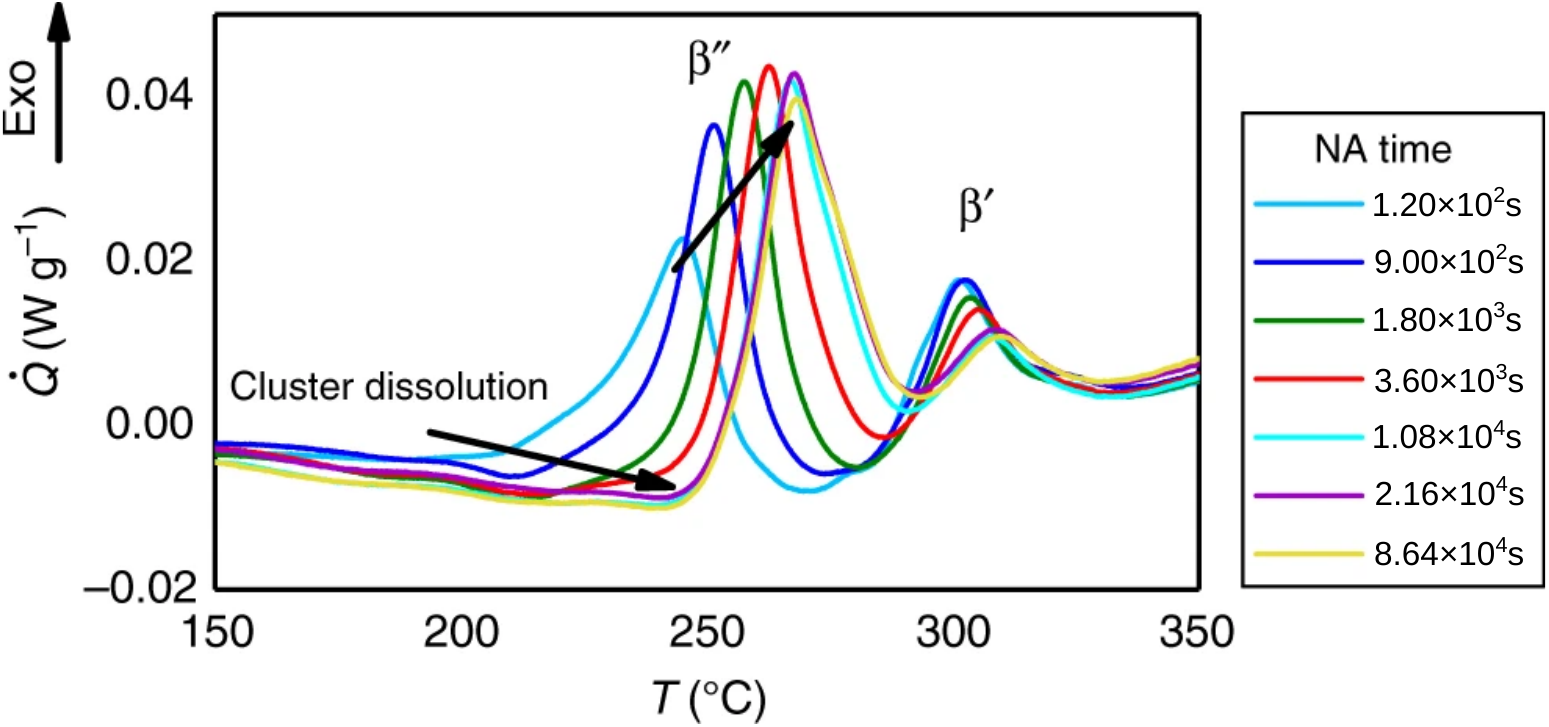}
    }\\
	\subfloat[Cluster number density vs size \cite{Zandbergen20151}]{\label{fig:c}\includegraphics[width=0.55\textwidth]{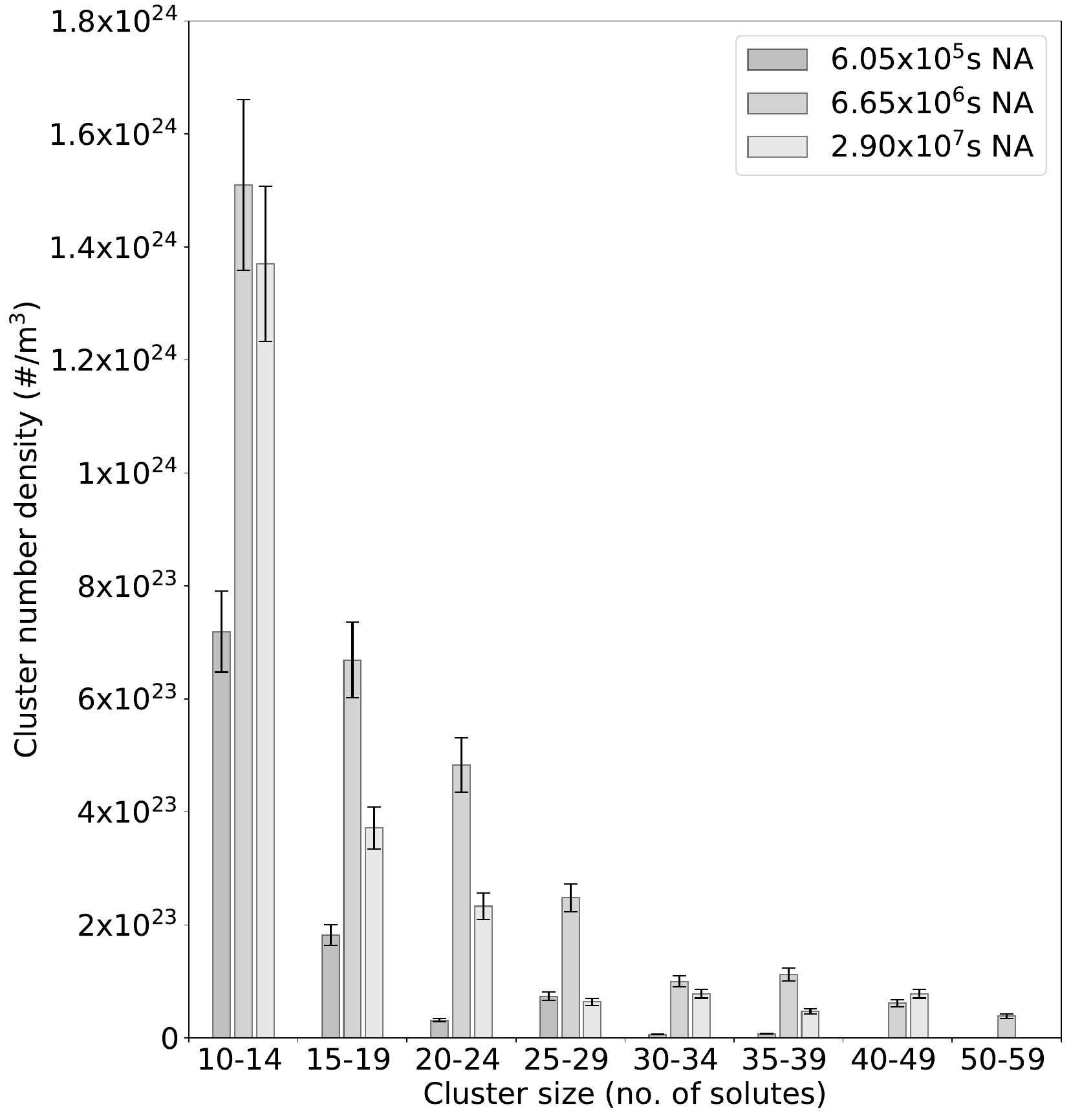}
    }\qquad%
	\subfloat[Clusters and Mg/Si ratio \cite{Zandbergen20152}. ]{\label{fig:d}\includegraphics[width=0.35\textwidth]{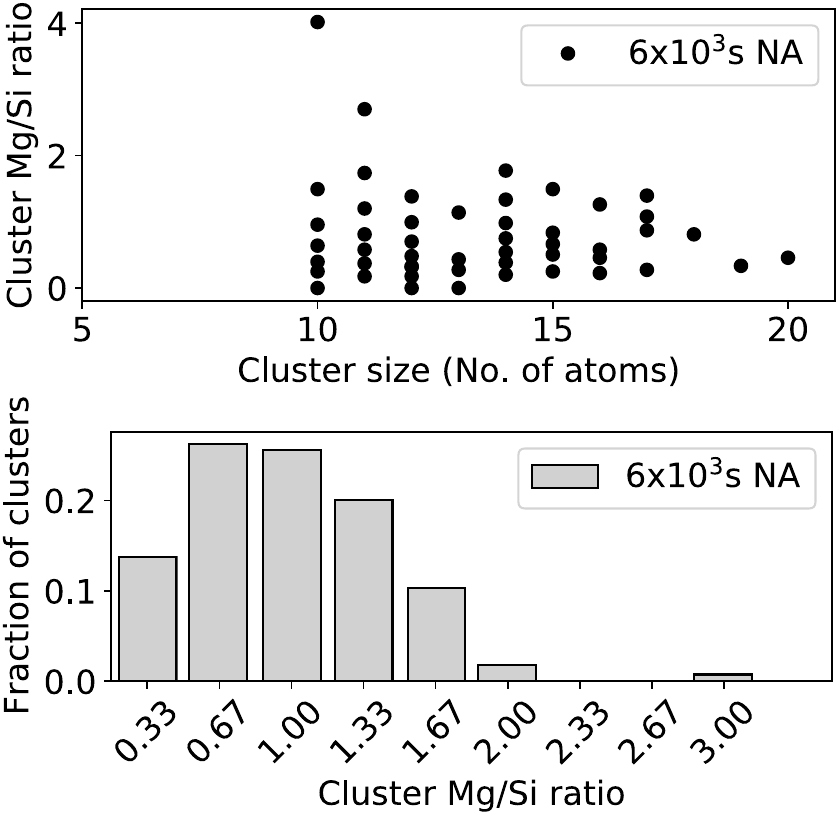}
    }%
	\caption{Experimental results on clustering during natural aging in Al-6xxx alloys.  (a) Hardness curves of 6061 Al alloy for artificial aging (AA) at 443 K and natural aging (NA) at 298 K, from the work of Pogatscher $et$ $al.$ \cite{Pogatscher2011}; (b) Excess heat flow (DSC heating curves) at varying natural aging times after quenching of an Al-Mg-Si alloy, reproduced from the work of Dumitraschkewitz $et$ $al.$ \cite{Dumitraschkewitz2019Size-dependentAlloys}; (c) Cluster number densities for different cluster size ranges, reproduced from Zandbergen $et$ $al.$ \cite{Zandbergen20152}; (d) Cluster Mg/Si ratio plotted against cluster size and fraction of clusters plotted against cluster Mg/Si ratio ranges after $6\times10^3$ s natural aging, reproduced from the work of Zandbergen $et$ $al.$ \cite{Zandbergen20151}. }
	
	\label{fig:myfig}
\end{figure}

In common industrial practice, there can be a considerable time between the initial alloy creation (steps (i) and (ii)) and the final artificial aging (step (iii)).  In some automotive applications, the artificial aging is performed during the ''paint-bake" cycle of manufacturing, long after the base alloy has been fabricated.  Unfortunately, Al-6xxx undergoes natural aging (NA) at room temperature.  Hence, the strength of the alloy when used after initial creation (steps (i) and (ii)) varies with time.  At the atomic scale, this aging is vacancy-mediated solute atom clustering that can occur at room temperature due to the excess quenched-in vacancy concentration \cite{Yang2021}.  Moreover, the solute atom clusters formed during natural aging are undesirable (see Figure \ref{fig:myfig}a).  First, their strengthening is insufficient for applications, compared to strengths attained by artificial aging \cite{Banhart2010}. Second, natural-aged samples show lower peak hardness after AA.  Finally, natural aging delays artificial aging, i.e. much longer AA times are needed to reach the maximum strengthening (note the logarithmic time scale in Figure \ref{fig:myfig}a).

The general explanation for the delay in AA after a long period of NA is that the natural aging clusters are not related to, nor can they directly transform to, the desired $\beta''$ precipitate or its precursor clusters.  That the NA clusters must be dissolved prior to starting an artificial aging sequence \cite{Murayama1998} is revealed in measurements of the heat flow during AA after different periods of NA, as shown in Figure \ref{fig:myfig}b.  Heat is first absorbed to dissolve the NA clusters, followed by heat release during formation of the $\beta''$ structures.  Increased natural aging times are associated with longer cluster dissolution times during AA.

Research into the nature of the NA clusters reveals common patterns of early stage clustering kinetics and size ranges but their structure(s) and composition(s) are still an issue of debate \cite{Dumitraschkewitz2019,Dumitraschkewitz2018,Geuser2006,Murayama1999,Edwards1998}.  Zanderbergen $et$  $al. $\cite{Zandbergen20151,Zandbergen20152} reported rapid clustering within $3.6\times10^5$ s of natural aging, after which the number density continues to increase for one week but the process slows down, as shown in Figure \ref{fig:myfig}c. The majority of clusters are between 10-15 atoms in size with Mg:Si ratios close to 1.0 (Figure \ref{fig:myfig}d), although 10 atoms was the minimum size considered in the analysis.   Analysis also suggests that the majority of solute atoms are still in solid solution or in very small clusters.  Similarly, Aruga $et$ $al.$ \cite{Aruga2015} also obtained increasing number density of clusters with natural aging time but only for the first ~$10^6$ s of aging, after which clustering was significantly slowed. Also, Mg:Si ratios in the range $0.8-1.0$ had the highest number density of clusters, consistent with results of Zanderbergen $et$ $al.$ \cite{Zandbergen20151,Zandbergen20152}. Marceau $et$ $al.$ \cite{Marceau2013} carried out Atom Probe Tomography (APT) on naturally-aged Al-6111 and reported increasing number density of solute atom clusters up to ~$6\times10^5$ s of natural aging time followed by near stagnation and the majority of Mg-Si clusters being in the range of 3-10 atoms. 

Based on the various experimental studies, we can summarize some common observations: (a) rapid clustering occurs within the first $\approx 10^4$ s of natural aging, (b) cluster density increases over the next $10^5-10^6$ s after which there is very little increase; (c) upon stagnation of growth, cluster sizes are typically around 10 atoms and the Mg/Si ratio in the clusters is close to 1.0; (d) artificial aging of naturally aged alloys first requires dissolution of clusters in order to precipitate $\beta''$.  A proposed explanation for the limited cluster growth after some period of natural aging is the trapping of vacancies by the clusters, i.e. formation of so-called 'vacancy prisons' \cite{Pogatscher2011}.  Vacancy trapping significantly reduces solute atom diffusion and, hence, further precipitate growth . However, it is challenging to verify this phenomenon based on experiments alone, and quantification requires accurate atomistic details. It is also unclear why the NA clusters that form are not precursors to $\beta''$ or cannot easily transform into the $\beta''$ precipitate without dissolution.  Existing computational studies of aging \cite{Myhr2001,Sha2005,Ye2021} have been limited in scope due to a lack of quantitative interatomic potentials that can accurately model precipitation energetics and kinetics.  This is not surprising since the development of chemically-accurate potentials for multi-component alloys has proven to be a serious challenge.  Recent advances in the creation of machine learning interatomic potentials (MLIPs) have now opened the possibility of examining the early states of precipitation at a level of accuracy close to that of Density Functional Theory (DFT) \cite{Jain2021}.  MLIPs can now be used to provide deeper insights into metallurgically-relevant phenomena.  

To provide qualitative and quantitative insight into the open issues regarding NA, we use a recently-developed Neural Network Potential (NNP) for the Al-Mg-Si \cite{Jain2021} in combination with off-lattice kinetic Monte Carlo (kMC) simulations at room temperature ($T=300$ K) to study the early stages of NA in a typical Al-6xxx alloy (see Methods).  Through our investigations, we are able to rationalize many of the aforementioned experimental observations.  Specifically, our study reveals that compact Mg-Si solute clusters of up to 15 atoms form within an estimated $5\times10^4$ seconds, consistent with APT findings. These clusters have a much lower formation energy than comparably-sized clusters that form an apparent fcc sub-motif for the $\beta''$ precipitate, and differ structurally.  However, we find that clusters of 14-18 atoms undergo the spontaneous formation of vacancy-Mg interstitial pairs that creates $\beta''$-like motifs, indicating a path for $\beta''$ formation.  Nonetheless, Si atoms replacing Mg in specific $\beta''$ sites still strongly stabilize non-$\beta''$ clusters, and the ''dissolution" to form $\beta''$ may be associated with replacing these Si atoms with the Mg atoms.  Furthermore, at sizes starting around 9-10 solute atoms, these solute clusters trap vacancies, with trapping energies of ~0.3 eV on surface cluster sites and ~0.5 eV on interior cluster sites.  This trapping is consistent with the significant stagnation of cluster density evolution observed in NA. Our results thus clarify many atomistic aspects of the undesirable NA in Al-6xxx alloys, consistent with experiments.

The remainder of this paper is organized as follows. Section 2 presents a preliminary analysis of Mg-Si clusters and their possible evolution, based on preliminary kMC results. Section 3 describes the kMC results and broad analysis of energetics and kinetics.  Section 4 discusses vacancy traps further.  Section 5 summarizes our main results.  The methods used in this study are presented in Section 6.

\section{Mg-Si cluster energetics} \label{sct:KMC}

{Before proceeding with kMC, it is valuable to examine the energetics of the likely precursor structures of the $\beta''$ phase.  A unit cell of the Mg$_5$Si$_6$ $\beta''$ structure, viewed along the $\bf{b}$ axis aligned with the Al [001] axis, is shown in Figure ~\ref{fig:betamotifs}a.  The unit cell consists of two Mg$_5$Si$_6$ formula units (f.u.), with a formation energy of -2.636 eV/f.u. (DFT).  Two other closely-related $\beta''$ phases are (i) the Mg$_5$Al$_2$Si$_4$ structure where the two outer Si atoms in each f.u. are replaced by Al atoms and (ii) the Mg$_4$Al$_3$Si$_4$ structure where the central Mg atom in each f.u. is replaced by an Al atom.  There is no $\beta''$ structure in which Si replaces the central Mg nor in which the square of 4 Si atoms has any substitution of Mg or Al.}

{ Within each formula unit of the Mg$_5$Si$_6$ structure in Figure \ref{fig:betamotifs}a, we can identify an Mg$_5$Si$_4$ cluster that appears close to 9-atom fcc structure (see Figure \ref{fig:betamotifs}b), with Si near the cube corners and Mg near the face centers.  When embedded coherently in an Al fcc lattice, this fcc 9-atom cluster has a formation energy of $-0.249$ eV.  In comparison, two other fcc 9-atom clusters with Mg on the cube corners and Si on the cube faces have much lower, and nearly-identical, DFT formation energies of $-0.463$ eV and $-0.464$ eV, respectively.  These two clusters differ only in the identity (Mg or Si) of the face atom in the center of the Mg cube corner atoms (see Figure \ref{fig:betamotifs}c,d).  Thus, the structure in Figure \ref{fig:betamotifs}b is not a viable $\beta''$ precursor structure.

\begin{figure}[H]
	\centering
	\includegraphics[width=4.5in]{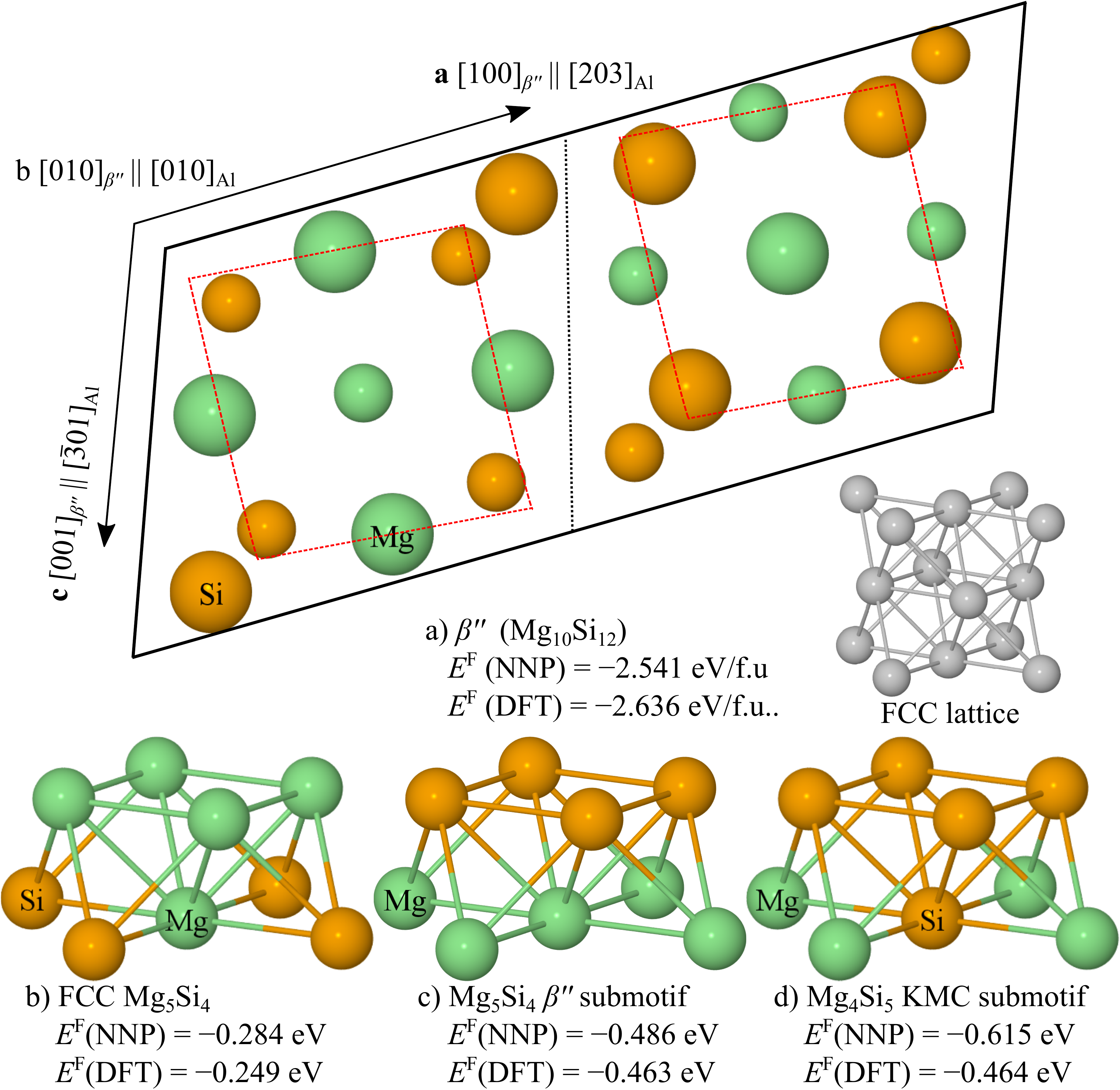}
	\caption{(a) $\text{Mg}_{10}\text{Si}_{12}$ $\beta''$ precipitate unit cell and lattice vectors, consisting of two 11-atom $\text{Mg}_{5}\text{Si}_{6}$ formula units (orange: Si; green: Mg; smaller atoms are in one atomic plane normal to the page and larger atoms are in a second atomic plane); (b) each formula unit has an apparent 9-atom $\text{Mg}_{5}\text{Si}_{4}$ near-fcc cluster as indicated by the orange squares in (a); (c) another 9-atom fcc $\text{Mg}_{4}\text{Si}_{5}$ pre-$\beta''$ cluster has been proposed \cite{Derlet2002}; but (d) there is an energetically-comparable 9-atom fcc $\text{Mg}_{5}\text{Si}_{4}$ cluster that is not easily related to $\beta''$. }	
	\label{fig:betamotifs}
\end{figure}

The Mg$_5$Si$_4$ cluster in Figure \ref{fig:betamotifs}c has previously been suggested as a pre-$\beta''$ structure \cite{Derlet2002}.  This pre-$\beta''$ structure is proposed to convert to the $\beta''$ structure by motion of the central Mg into an interstitial position in between the Si atoms, leaving a vacancy behind.  This is consistent with the understanding that the $\beta''$ formula units can be viewed as fcc-like but with a central column of Mg atoms uniformly shifted by b/2 along the $\bf{b}$ axis, forming an alternating array of Mg interstitials and vacancies.  However, why and how such a shift occurs has not been established, and formation of vacancy/interstitial pairs is usually thought to be energetically prohibitive.  In addition, the Mg$_4$Si$_5$ in Figure \ref{fig:betamotifs}c having the central Si atom has $\emph{not}$ been considered previously because it is not consistent with the $\beta''$ site occupancies.  Yet it has nearly the same formation energy as the pre-$\beta''$ Mg$_5$Si$_4$ cluster.  The NNP-computed formation energies for these two competing clusters agree reasonably well with DFT but the Mg$_4$Si$_5$ cluster is predicted to be $~0.15$ eV more stable than found via DFT.}

\begin{figure}[H]
	\centering
	\includegraphics[width=6.5in]{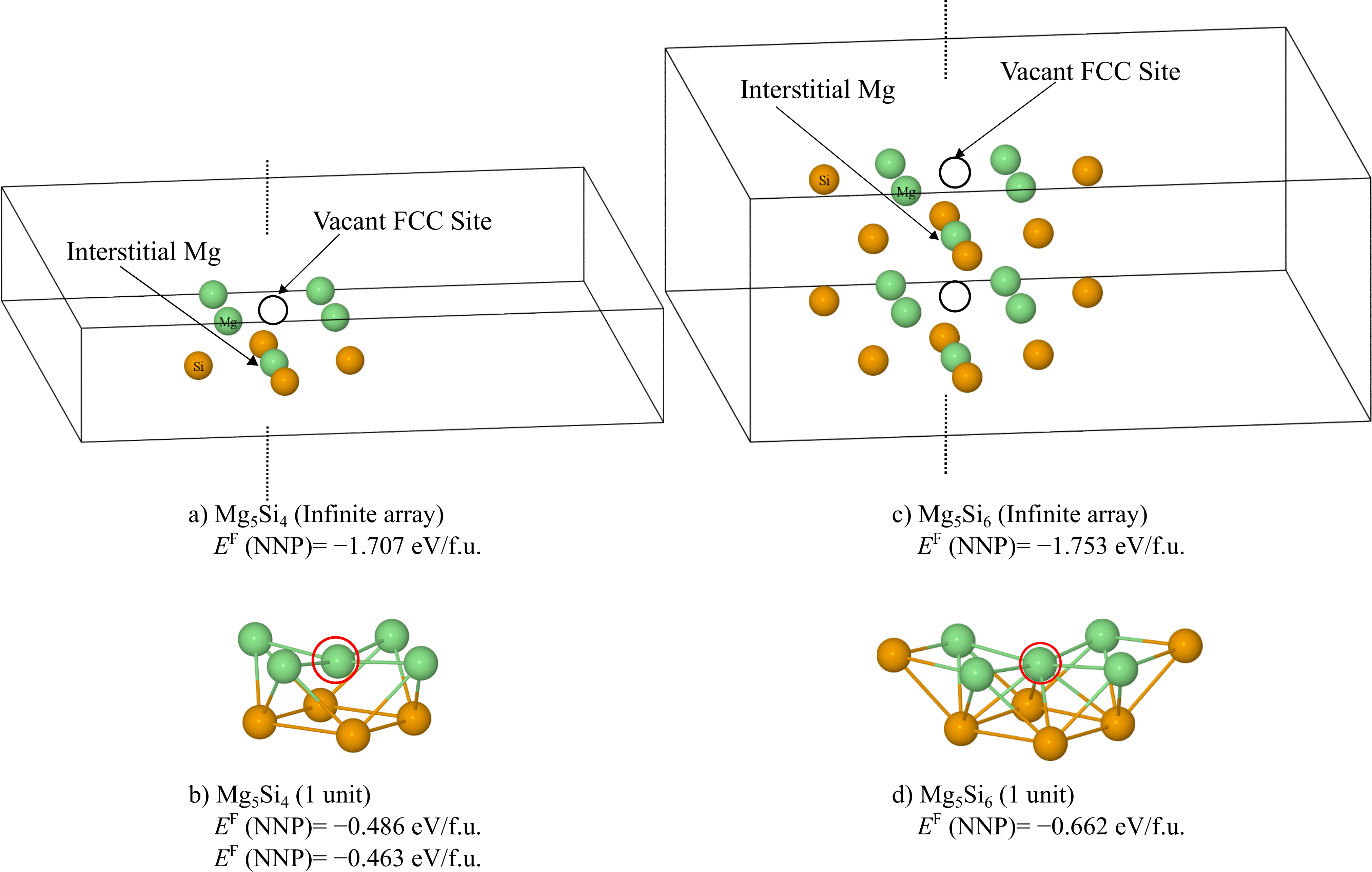}
	\caption{(a) Periodic array of stacked $\text{Mg}_{5}\text{Si}_{4}$ $\beta''$ 9-atom formula units with a vacancy-Mg interstitial pair, as indicated, rather than a structure fully coherent with the fcc lattice; (b) a single sub-formula unit of $\text{Mg}_{5}\text{Si}_{4}$ $\beta''$ coherent with the fcc lattice due to the relaxation of the Mg (circled in red) back to the fcc lattice position; (c) periodic array of the $\text{Mg}_{5}\text{Si}_{6}$ $\beta''$ formula unit with a vacancy-Mg interstitial pair, as indicated, rather than a structure fully coherent with the fcc lattice; (d) a single formula unit of $\text{Mg}_{5}\text{Si}_{6}$ $\beta''$ coherent with the fcc lattice due to the relaxation of the Mg (circled in red) seen in (c). Formation energies for all structures as computed by DFT and/or NNP are shown, both total and per 9-atom formula unit.}	
	\label{fig:fig3}
\end{figure}

{The presence of the Mg-interstitial/vacancy columns in $\beta''$ suggests another $\emph{non-fcc}$ $\text{Mg}_{5}\text{Si}_{4}$ cluster in which the Mg atoms are on the fcc cube corners, the Si atoms are on the fcc faces, and the central Mg moves to an interstitial position in the center of the 4-atom Si face atoms, leaving a vacancy in the center of the 4-atom Mg cube corner atoms.  As proposed earlier ~\cite{Derlet2002}, this is formed by moving the central Mg in the pre-$\beta''$ structure of $\text{Mg}_{5}\text{Si}_{4}$ into the interstitial position. Creating an infinite periodic needle of this configuration with the infinite column of alternating Mg-interstitial/vacancy found in $\beta''$ (Figure \ref{fig:fig3}a), the NNP predicts a stable structure with a very low formation energy of $-1.707$ eV per 9-atom f.u. However, when a single such $\text{Mg}_{5}\text{Si}_{4}$ cluster is embedded in the Al matrix, it is $\emph{not}$ stable (in both DFT and NNP).  It spontaneously transforms back to the $\text{Mg}_{5}\text{Si}_{4}$ structure of Figure \ref{fig:betamotifs}c and shown again in Figure \ref{fig:fig3}b}.

{The above study neglects the additional 2 Si atoms present in the Mg$_5$Si$_6$ formula unit (Figure ~\ref{fig:betamotifs}a), and so is more closely related to Mg$_5$Al$_2$Si$_4$.  However, the additional 2 Si atoms do not influence the general process above.  Specifically, as shown in Figure ~\ref{fig:fig3}c, a periodic needle of $\text{Mg}_{5}\text{Si}_{6}$ has a formation energy of -1.753 eV/f.u. that is only slightly lower than the -1.707 eV/f.u. for the $\text{Mg}_{5}\text{Si}_{4}$ cluster.  Furthermore, a single $\text{Mg}_{5}\text{Si}_{6}$ cluster embedded in the Al matrix shows the central Mg atom to be stable in the fcc position (Figure ~\ref{fig:fig3}d), as found for the $\text{Mg}_{5}\text{Si}_{4}$ cluster, and having a formation energy of -0.662 eV/f.u. only slightly lower than that for the $\text{Mg}_{5}\text{Si}_{4}$ cluster.  The addition of the two Si atoms to form the $\text{Mg}_{5}\text{Si}_{6}$ cluster does reduce the formation energy slightly, but does not alter the general path found for the $\text{Mg}_{5}\text{Si}_{4}$ clusters.}

\begin{figure}[H]
	\centering
	\includegraphics[width=5in]{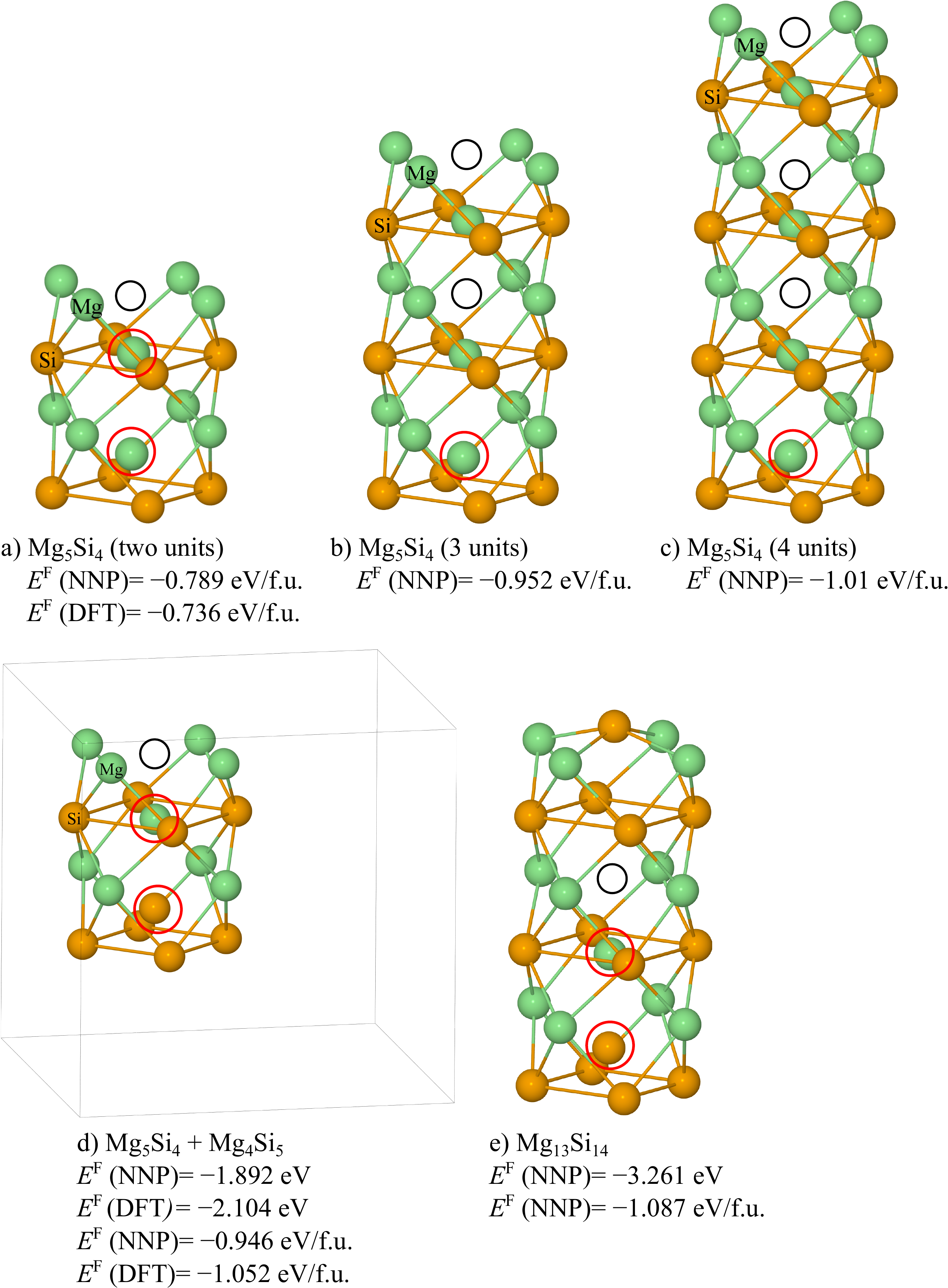}
	\caption{a,b,c) Stacks of two, three, and four $\text{Mg}_{5}\text{Si}_{4}$ $\beta''$ 9-atom formula units, showing the formation of one, two, and three vacancy-Mg interstitial pairs, respectively, with one partial interstitial pair (Mg circled in red);  (d) 18-atom cluster stack of Mg$_4$Si$_5/$Mg$_5$Si$_4$ $\beta''$ where the bottom center Mg atom in (a) has been replaced by Si, resulting in a much lower formation energy;  (e) 27-atom cluster stack of Mg$_4$Si$_5/$Mg$_5$Si$_4/$Mg$_4$Si$_5$ where both top and bottom center Mg atoms in (b) have been replaced by Si, resulting in a much lower formation energy.  In both (d) and (e), the remaining central Mg atom moves to the interstitial position.  Formation energies for all structures as computed by DFT and/or NNP are shown, both total and per 9-atom formula unit.
    }
	\label{fig:fig4}
\end{figure}

{To understand evolution from the pre-$\beta''$ (Figure \ref{fig:fig3}b)  to the infinite periodic array (Figure \ref{fig:fig3}a), we examine the 18-atom cluster formed by stacking two $\text{Mg}_{5}\text{Si}_{4}$ units (Figure \ref{fig:fig4}a). The low-energy configuration of this structure shows the spontaneous formation of one vacancy-Mg interstitial pair and relaxation of the second internal Mg into an intermediate interstitial position, with a formation energy of $-0.736$ eV/f.u. (NNP energy -$0.789$ eV/f.u.).  The 27- and 36-atom clusters formed by stacking 3 and 4 $\text{Mg}_{5}\text{Si}_{4}$ units, respectively (Figures \ref{fig:fig4}b,c), embedded in longer simulation FCC $3\times3\times6$ supercells - double the height along the stacking direction), reveal that the vacancy-Mg interstitial is formed in 2 and 3 of the units, respectively, but with the terminal unit continuing to have only a partially-formed vacancy-Mg interstitial pair.  The formation energies continue to decrease, to $-0.952$ eV/f.u. (NNP) and $-1.01$ eV/f.u. (NNP), respectively. This indicates a slow convergence toward a steady reduction in energy of 1.2-1.4 eV with successive addition of $\text{Mg}_{5}\text{Si}_{4}$ units.  The difference between the infinite periodic needle ($-1.707$ eV) and the converging energy ($-1.4$ eV) of the evolving stack of $\text{Mg}_{5}\text{Si}_{4}$ units is presumably attributable to the surface energies between the cluster atoms and the Al matrix at the top and bottom of the evolving stack.  These results suggest that the $\beta''$ structure forms starting from needle growth of the pre-$\beta''$ $\text{Mg}_{5}\text{Si}_{4}$ structure, with a natural spontaneous formation of the vacancy-Mg interstitial pair that is an essential feature of the $\beta''$ structure. }

{Having shown that the $\text{Mg}_{4}\text{Si}_{5}$ cluster (Figure \ref{fig:betamotifs}d) is energetically competitive with the $\text{Mg}_{5}\text{Si}_{4}$ cluster (Figure \ref{fig:betamotifs}c), it is important to examine further the energetic consequences of the substitution of Si into the central Mg site.  As shown in Figure~\ref{fig:fig4}d, an 18-atom cluster formed by stacking one $\text{Mg}_{5}\text{Si}_{4}$ unit on top of one $\text{Mg}_{4}\text{Si}_{5}$ unit leads to a much lower formation energy of $-2.104$ eV ($-1.892$ eV for the NNP) as compared to the 18-atom cluster formed by the stack of two $\text{Mg}_{5}\text{Si}_{4}$ units ($-1.472$ eV with DFT; $-1.578$ eV with NNP).  The presence of just the one Si atom in the bottom unit has a distinct and significant ($-0.632$ eV in DFT, $-0.314$ eV in NNP) stabilizing effect on a structure that is $\emph{not}$ consistent with the $\beta''$ structure.  Furthermore, a 27-atom cluster consisting of a stacking of $\text{Mg}_{4}\text{Si}_{5}$/$\text{Mg}_{5}\text{Si}_{4}$/$\text{Mg}_{4}\text{Si}_{5}$ where Si replaces the central Mg in both end units (~\ref{fig:fig4}e),has an even lower energy ($-3.261$ eV for NNP) compared with the 27-atom $\beta''$ structure in Figure \ref{fig:fig4}b ($-2.856$ eV for NNP).  These more-stable non-$\beta''$ clusters should thus form preferentially over the $\beta''$ clusters, preventing $\beta''$ formation.  These more-stable non-$\beta''$ clusters can only be converted to $\beta''$ clusters by the replacement of the terminal Si atom(s) by Mg atom(s) at a significant energy cost.  This replacement of Si by Mg may be the ''dissolution" process (requiring heat input; see Figure \ref{fig:myfig}b) that must occur to enable the formation of $\beta''$ during artificial aging.  This is an important finding of the present study.}

\begin{figure}[H]
	\centering
	\includegraphics[width=6.5in]{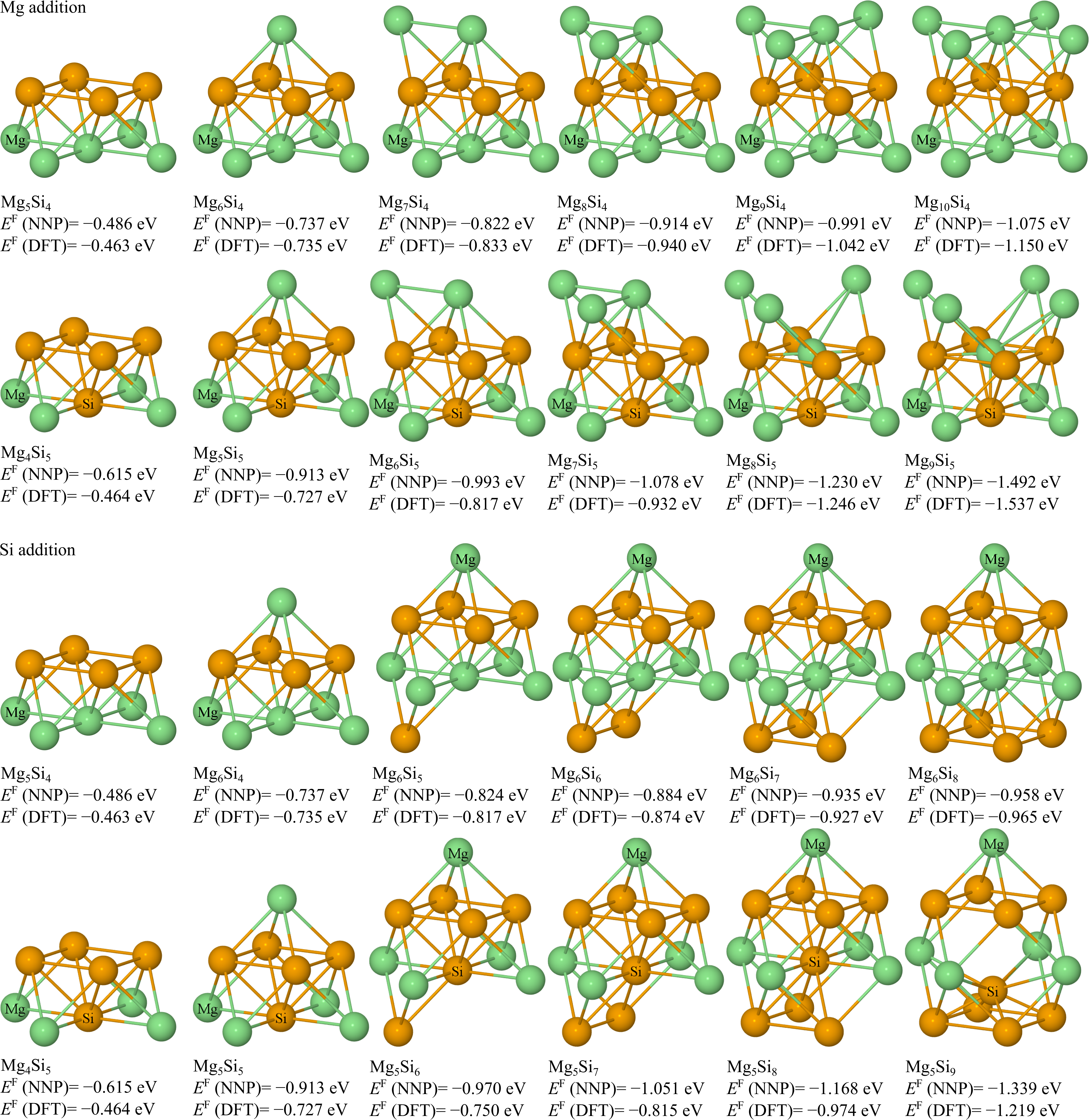}
	\caption{Formation energies for evolving solute atom clusters upon sequential Mg and Si additions as shown, starting from the 9-atom Mg$_5$Si$_4$ and Mg$_4$Si$_5$ $\beta''$ clusters. The first Mg addition leads to a significant increase in formation energy, therefore the Si additions are considered only after the first Mg addition. Only the Mg$_8$Si$_5$, Mg$_9$Si$_5$ and Mg$_5$Si$_9$ structures show a transformation to form a vacancy-Mg interstitial with a significant increase in the formation energy.  
	}
	\label{fig:fig5}
\end{figure}

The above studies involve stacks of the pre-$\beta''$-type structures on an fcc lattice, and so are limited to 9, 18, .... atom clusters.  It is thus useful to examine the energetics of compact clusters starting from the 9-atom clusters on the fcc lattice and adding sequential Mg or Si atoms to evolve toward the two-unit 18-atom structures examined above.  Figure~\ref{fig:fig5} shows the formation energies of clusters formed by the addition of Mg atoms to the initial Mg$_5$Si$_4$ and Mg$_4$Si$_5$ clusters, respectively. The first notable feature is that the addition of a single Mg above the 4-atom cube-face Si atoms leads to a large reduction (approximately $-0.27$ eV) of the cluster formation energy.  The 10-atom Mg$_6$Si$_4$ and Mg$_5$Si$_5$ clusters that are particularly stable against dissolution as compared to smaller clusters.  

Adding further Mg atoms to the Mg$_6$Si$_4$ cluster leads to a steady modest reduction in formation energy, with the fcc structure retained up to the 14-atom Mg$_{10}$Si$_4$ cluster.  In the 14-atom Mg$_{10}$Si$_4$ cluster there is no indication that either of the face-centered Mg atoms are unstable to the formation of a vacancy-Mg interstitial pair seen in Figure ~\ref{fig:fig4}.  Evidently, the further additional 4 Si atoms in the Mg$_{10}$Si$_8$ cluster are needed to drive this transformation toward the $\beta''$ structure.   Adding further Mg atoms to the  Mg$_5$Si$_5$ cluster leads to a similar decrease in energy for the first few Mg additions. However, the 13-atom cluster  Mg$_8$Si$_5$ shows the spontaneous formation of the vacancy-Mg interstitial structure with a concomitant significant lowering ($-0.3$ eV) of the formation energy, as compared to the change of $-0.1$ eV between  Mg$_8$Si$_4$ and  Mg$_9$Si$_4$.  The addition of a final Mg to form the Mg$_9$Si$_5$ cluster leads to a further large decrease ($-0.29$ eV).  Thus, the presence of just the one additional Si atom in place of the Mg atom has very little effect on the cluster energetics up to Mg$_8$Si$_4$ and Mg$_7$Si$_5$ but then drives a significant change in both structure and energy upon further Mg addition.  

For both  Mg$_8$Si$_5$ and Mg$_9$Si$_5$, the NNP predicts that the DFT structure is stable with a similar energy as shown. $\it{However}$, the NNP also predicts a $\it{metastable}$ structure coherent with the fcc lattice, i.e. without the vacancy-Mg interstitial, with only slightly higher energies ($-1.159$ eV and $-1.239$ eV, respectively).  The NNP thus has a very small non-zero barrier that prevents the spontaneous formation of the $V$-Mg interstitial pair.  kMC studies shown later, which are executed assuming an underlying fcc lattice, may therefore not show the formation of the vacancy-Mg interstitial structures in these 13- and 14-atom clusters.  Since formation of vacancy-Mg interstitial structures violates the assumptions of the kMC model, this error in the NNP is useful but must be recognized as influencing interpretations of kMC for these larger clusters. 

Figure~\ref{fig:fig5} also shows the formation energies of clusters formed by the addition of Si atoms to Mg$_6$Si$_4$ and  Mg$_5$Si$_5$ (after the addition of the one additional Mg).  The Si additions lead to a steady lowering of the formation energy in both cases.  The DFT results for the two sets of clusters are also very similar up to 13-atoms, indicating the presence of the additional Si in place of the Mg is not energetically important up to this point.  However, upon addition of one final Si atom, the central Si atom in the Mg$_5$Si$_9$ cluster spontaneously moves to form a vacancy-Si interstitial pair but with the Si remaining above the plane of the 4 Si atoms beneath it.  No such structural change is observed for the Mg$_6$Si$_8$ cluster, where the Mg replaces the central Si atom.  

The results in Figure~\ref{fig:fig5} show (i) a notable stability of adding one Mg on top of 4 Si face-centered atoms to form a 10-atom cluster, (ii) a steady increase in the formation energy (more negative) indicating no metastable clusters that would inhibit continued growth of the clusters, and (iii) the role of one face-centered Si atom in driving a structural transition in 13-14 atom clusters accompanied by a significant reduction in energy in spite of the formation of vacancy-Mg or vacancy-Si interstitial pairs.  While the 18-atom clusters with both the Mg and Si face centered atoms are found to enable the formation of the vacancy-Mg interstitial pair, this structural transformation occurs preferentially and at an earlier stage (smaller clusters of 13-14 atoms) in the presence of the Si face centered atom.  {However, as found for the 18- and 27-atom clusters, since the face-centered Si is not a feature of the $\beta''$ structure, these energetically-favorable clusters must ultimately be transformed into higher-energy structures by replacement of the Si atom for a Mg atom before the $\beta''$ structure can be formed and grow.}

In all cases studied above, we note that any structural transformation appears to require 13-18 atom clusters, depending on the precise composition and arrangement.  Experimental observations suggest that many clusters are stabilized at 10-14 atoms (and these are not necessarily the very compact structures formed here), so that the initial clustering during natural aging may largely remain coherent with the underlying fcc Al lattice.
 
Finally, our results here include many comparisons between DFT and NNP energies.  Overall, the NNP results are in very good agreement with the DFT results across a wide range of compositions and cluster sizes and configurations.  The only notable deviations arise for the initial Mg$_4$Si$_5$ cluster and clusters derived from it by adding a few additional Si or Mg atoms.  The NNP was not trained on any of these clusters, and so we can anticipate that the NNP can provide the accuracy needed to proceed with kinetic studies of cluster evolution during natural aging.

\section{Natural aging} 

The sizes and configurations of the clusters studied in the previous section were motivated by preliminary results from kMC simulations of natural aging.  We now turn to a more systematic kMC study to examine the kinetics of cluster formation and the role of vacancies in controlling the kinetics.

\subsection{Descriptive overview of kMC results}

\begin{figure}[H]
	\centering
	\includegraphics[width=6.5in]{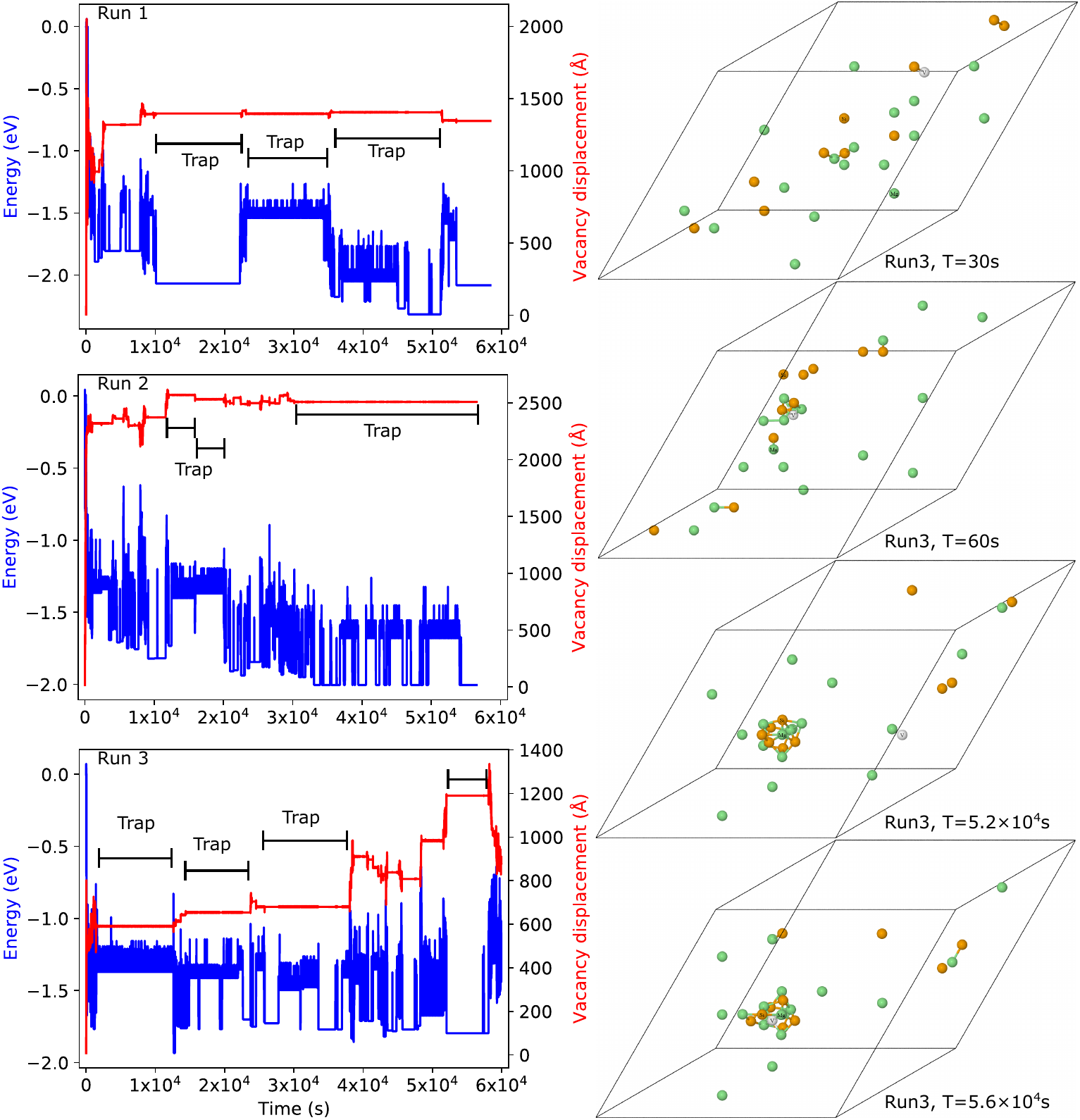}
	\caption{ Energy and RMS vacancy displacement vs simulation time for three different kMC simulations carried out with the NNP using a 1728 atoms supercell containing 17 Mg atoms (green), 10 Si atoms (orange), and 1 vacancy (pink).  Geometries have been extracted at four different times from the third trajectory. The simulations were carried out for $1.5\times10^7$ kMC steps. 
	}
	\label{fig:enemg20}
\end{figure}

Figure \ref{fig:enemg20} shows three examples of the energy versus (estimated rescaled) time and RMS vacancy displacement versus time for three kMC simulations of natural aging in Al-Mg-Si.  Atomistic configurations at different stages of clustering in the third trajectory are also shown. We first discuss the broad trends in a descriptive manner and later turn to quantification.  At short times ($\sim10^2$ s), the system energy decreases significantly, corresponding to the formation of small clusters ($\sim6$ atoms).  For instance, in the first geometry shown in Figure \ref{fig:enemg20}, there are a 4-atom and a 2-atom cluster at 30s.  In the second geometry shown a 6-atom cluster has formed at 60 s.  These clusters are too small to strongly trap the vacancy, and so the vacancy escapes quickly and continues to transport more solute atoms toward the cluster(s).  Larger clusters start forming at $\sim10^3$ s, consistent with experimental results which show incremental hardness increases at $\sim10^3$ s.  During this time, the vacancy diffusion distance is on the order of 1000-2000 \AA, consistent with recent experiments of the distances needed to form clusters \cite{Dumitraschkewitz2019}.  Once the energy has decreased to the range of $-1.0$ to $-1.5$ eV, one large cluster is typically formed that then varies in size (number of solute atoms) as solute atoms diffuse away from or into the cluster.  The total energy fluctuates as solute atoms are added or removed from the major cluster.  During cluster rearrangement, the changes in vacancy displacement are very small, indicating that the vacancy is trapped in/at the cluster.  During such periods, the system energy can fluctuate as the vacancy migrates in and around the cluster.  The vacancy motion changes the precise solute atom configuration and, hence, the cluster energy.  The vacancy can occasionally escape from the cluster, as illustrated in the third geometry in \ref{fig:enemg20}.  This is accompanied by a very rapid jump in the vacancy displacement and a notable increase in energy, the latter indicating that the vacancy had been bound to the cluster. The vacancy can then diffuse through the system, form and dissolve small clusters, and transport solute atoms to or from the larger cluster.  At some later time, the vacancy returns to the cluster and is trapped again, as illustrated in the fourth geometry in Figure \ref{fig:enemg20}.  The larger clusters generally persist throughout the time of the simulation so that large cluster dissolution was rarely observed. 

\subsection{Analyses of natural aging clusters} \label{sct:cluster}

{The kMC simulations regularly reveal the formation of the 9-atom pre-$\beta''$ Mg$_{5}$Si$_{4}$ and Mg$_{4}$Si$_5$ clusters (Figure~\ref{fig:betamotifs}c,d). The Mg$_{4}$Si$_5$ cluster is seen more frequently because the NNP formation energy is somewhat lower.  In reality, we can expect both types of clusters to arise with roughly equal probability.}

\begin{figure}[htbp]
	\centering	
	\includegraphics[width=3in]{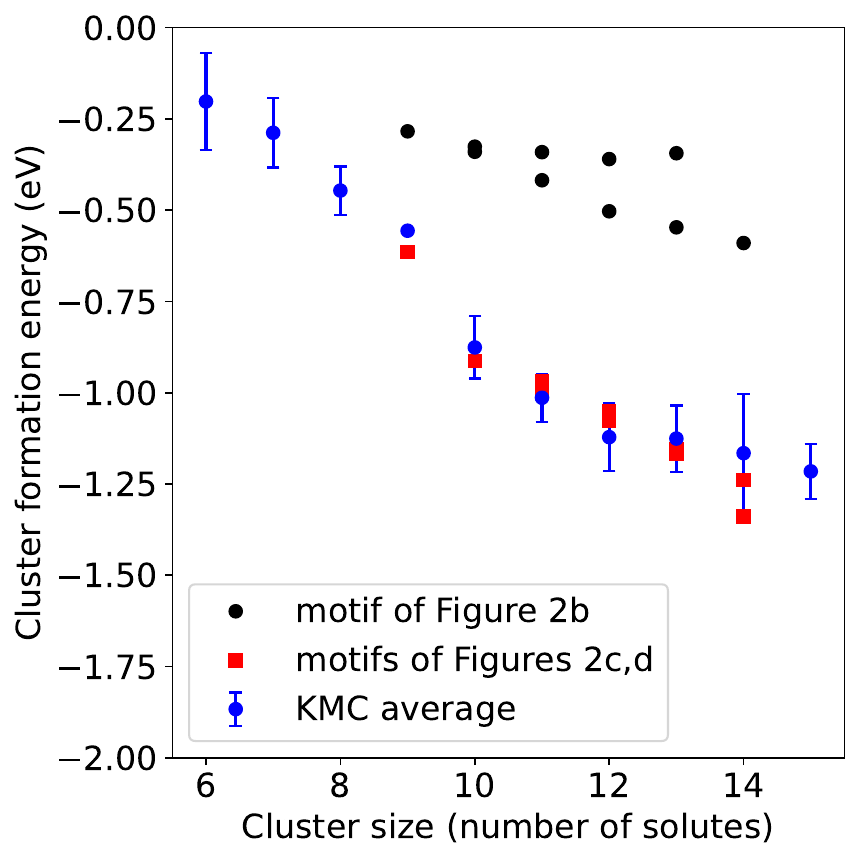}	
	\caption{Average formation energies of kMC clusters (blue) compared with the formation energies of clusters obtained from the motif of Figure \ref{fig:betamotifs}b (black), and the formation energies of clusters obtained from the motifs of Figures \ref{fig:betamotifs}c,d (red). }	
	\label{fig:myfig3}
\end{figure}

During the kMC, we determine the energy of the dominant cluster as a function of cluster size independent of cluster geometry.  A cluster is determined as a group of solute atoms all of which have at least one other solute atom in the cluster as a first nearest neighbor (near-neighbor).  As shown in Figure~\ref{fig:myfig3}, the cluster energies monotonically decrease with increasing cluster size, and closely follow the energies of the compact clusters shown in Figure~\ref{fig:fig5}.
The energies of clusters growing from the Mg$_5$Si$_4$ cluster (Figure~\ref{fig:betamotifs}b) are much higher, and these clusters are never observed in the kMC.  

Figure~\ref{fig:myfig3} shows a significant increase in cluster formation energy between 9 and 10 solute atoms.  This is presumably due to the very favorable addition of one Mg to either the Mg$_5$Si$_4$ or Mg$_4$Si$_5$ sub-motifs as shown earlier.  Overall, we can conclude more broadly than in the previous section that there are no thermodynamic barriers to growth of clusters up to sizes of 14-15 atoms.

The sizes of the clusters observed in the kMC are typically 8-14 atoms.  The largest clusters observed during the simulation contain 15 solute atoms. These results are consistent with experiments. The Mg:Si ratios in these clusters are also typically in the range 0.8-1.0, again consistent with experiments \cite{Aruga2015,Zandbergen2015}.  There is a slightly enhanced Si content partially due to the NNP bias toward clusters with the additional Si atom. Finally, the typical clusters formed do not consume all of the solute atoms in the system - roughly half of the solute atoms remain in solid solution throughout the kMC simulation time.  This is consistent with experiments indicating that a significant fraction of the Mg and Si atoms remain in solution, in spite of the existence of low-energy clusters.  This arises in part because the system is kinetically limited by the trapping of vacancies, as discussed next.

\section{Vacancy traps} \label{sct:traps}

The lack of cluster growth beyond 10-15 atoms suggests that such clusters are sufficiently large to trap vacancies for extended time periods and prevent further solute transport.  Once trapped, the vacancy may migrate within and adjacent to the cluster, but not escape and is hence unable to transport more solute atoms to the large cluster(s).  This is consistent with the kMC observations of long time spans over which the vacancy is not migrating away from the large atom cluster.  This expectation of local rearrangements is consistent with the results shown in Figure \ref{fig:enemg20}: in regions where the vacancy displacement is nearly constant, there can be large energy fluctuations caused by vacancy motion in/around the cluster. 

\begin{figure}[H]
	\centering
	\includegraphics[width=6in]{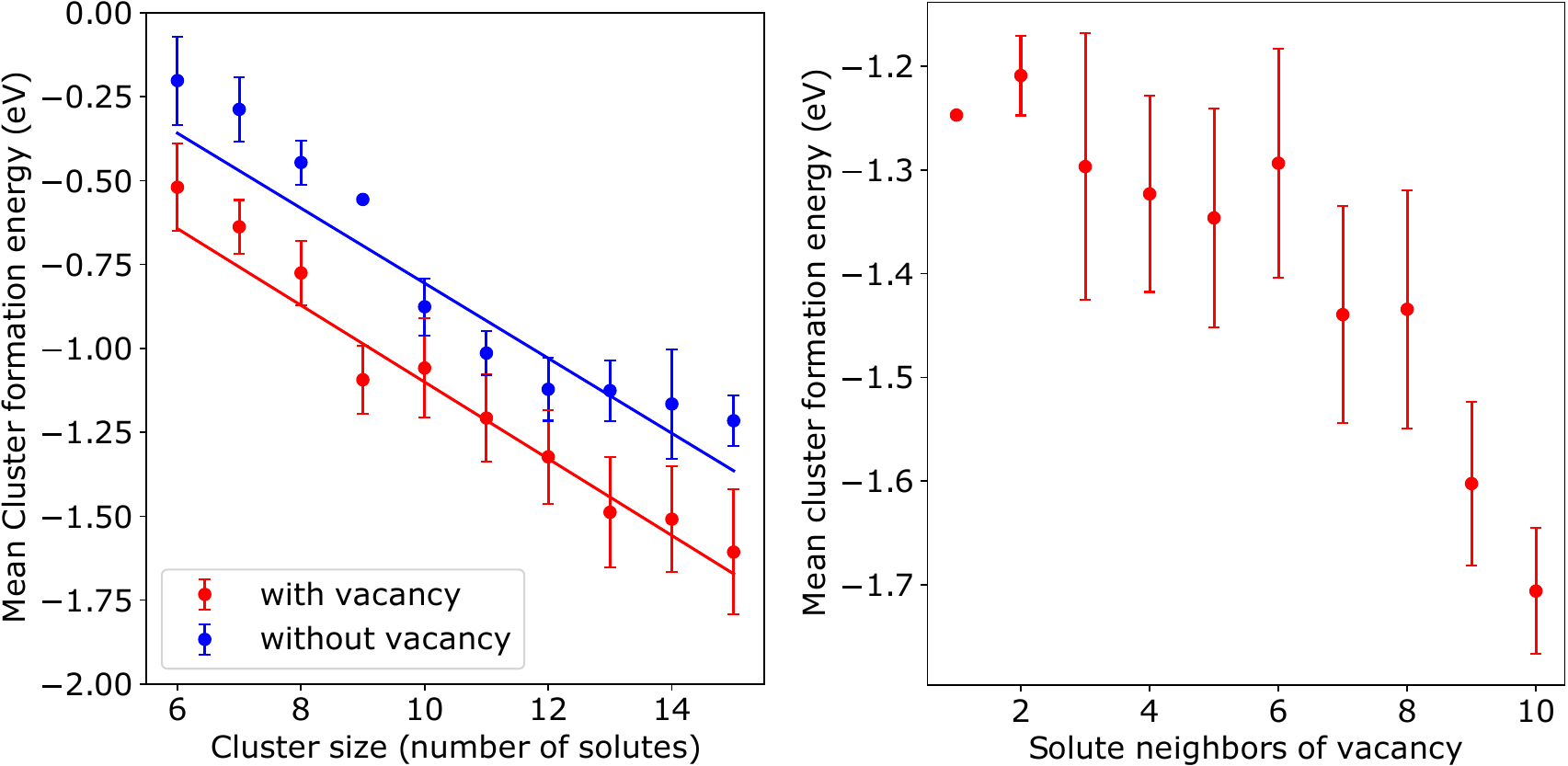}
	\caption{Left: Mean cluster formation energies vs cluster size for clusters with a vacancy (red) and without a vacancy (blue); Right: mean cluster formation energy vs number of vacancy near-neighbors for a fixed cluster size of 13 solute atoms. 
	}
	\label{fig:fig8}
\end{figure}

Comparing the energies of the same size clusters (same number of solute atoms) with and without a vacancy, as shown in Figure \ref{fig:fig8}a, reveals that the presence of the vacancy lowers the cluster energy on average by approximately 0.3 eV.  That is, the vacancy binding energy to the clusters, measured over many clusters and configurations, is typically 0.3 eV.  Note that the data in Figure~\ref{fig:fig8}a does not account for vacancy residency time, and so reflects configurations only.  So, most configurations have a binding energy of around 0.3 eV even though the deepest traps can have larger binding energies.  Examining the difference between the lowest energy clusters with and without a vacancy, differences of $~0.5$ eV are seen for larger cluster sizes, indicating that some vacancy traps have binding energies of $~$0.5 eV.  As discussed below, the smaller trapping energies tend to be associated with the vacancy occupying sites on the surface of the cluster, where there are 3 to 4 Si-vacancy neighbor pairs since Si-vacancy binding is favorable and Mg-vacancy binding is not.  The occasional larger trapping energies of 0.5 eV arise when the vacancy occupies sites in the center of the cluster, maximizing the number of Si-V near-neighbors.  

Figure \ref{fig:fig8}b further shows that the total formation energies of configurations with a trapped vacancy scales with the number of solute atom neighbors of the vacancy.  For a fixed cluster size of 13 solute atoms, the cluster formation energy is in the range of $~-1.3$ eV when the vacancy has 1-6 solute atom neighbors, suggesting the vacancy is on the surface of the cluster.  With increasing number of solute neighbors, i.e. as the vacancy becomes surrounded by solute atoms, the cluster formation energy including the vacancy decreases considerably, reaching $~-1.65$ eV, for 9-10 solute atom neighbors corresponding to a vacancy in the interior of the 13-atom solute clusters.  The lowest energy 13-atom clusters with a vacancy have formation energies ($-1.7$ eV) that are comparable to those of embedded $\beta''$ precipitates with 18 solute atoms $\emph{without}$ a vacancy (see Figure \ref{fig:fig4}).  Hence, these clusters with trapped vacancies are also preventing nucleation of larger $\beta''$ nuclei.   These low-energy vacancy-containing clusters are thus expected to persist for long times, which is seen in the direct kMC results as a stagnation in evolution of the system energy over long times.  The vacancy trapping is thus inhibiting solute atom transport and cluster growth kinetics, and significantly retarding the formation of $\beta''$.  

\begin{figure}[H]
	\centering
	\includegraphics[width=5in]{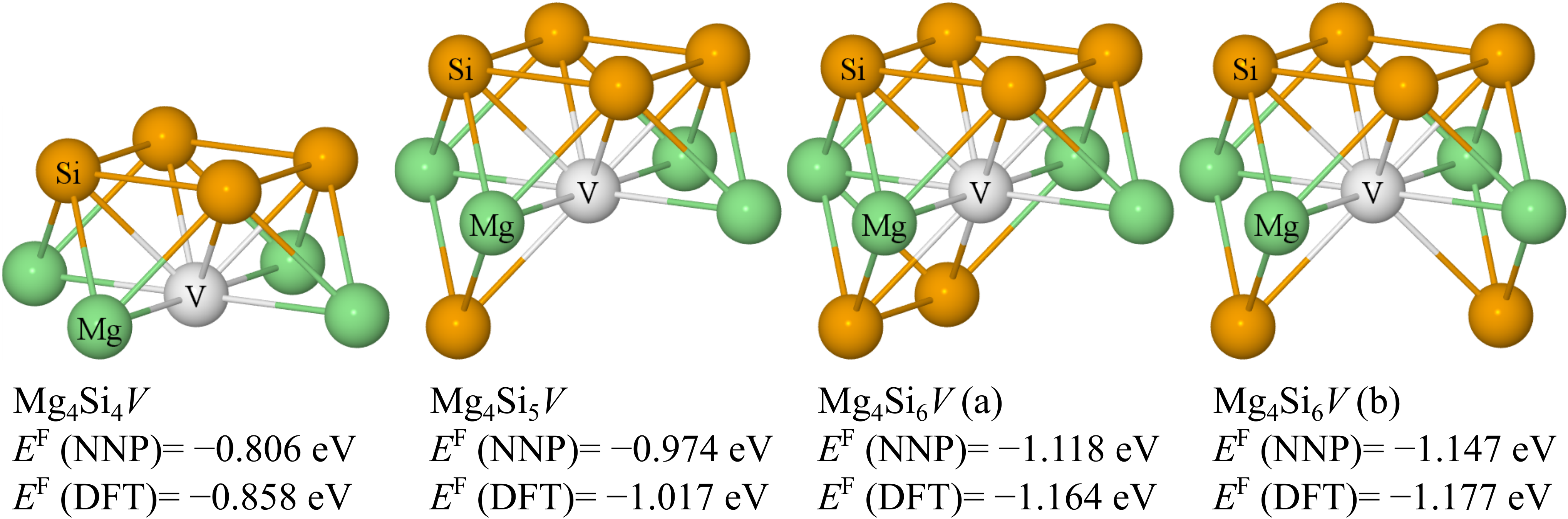}
	\caption{Configurations and formation energies of some typical low-energy vacancy traps for clusters of 8-10 solute atoms, as observed during kMC simulations. 
	}
	\label{fig:traps}
\end{figure}

We have examined some of the strongest vacancy trapping configurations arising in the kMC, and have compared their total energies to DFT as shown in Figure \ref{fig:traps}. The vacancy is typically in the center of the cluster, with as many solute atom neighbors as possible.  The cluster formation energy decreases as the number of Si solute atoms increases.  Comparisons with DFT show very good quantitative agreement, validating the good accuracy of the NNP for cluster energetics even in the presence of a vacancy.  

\begin{figure}[H]
	\centering
 \begin{minipage}[b]{.7\linewidth}
	\centering
	\includegraphics[width=5.5in]{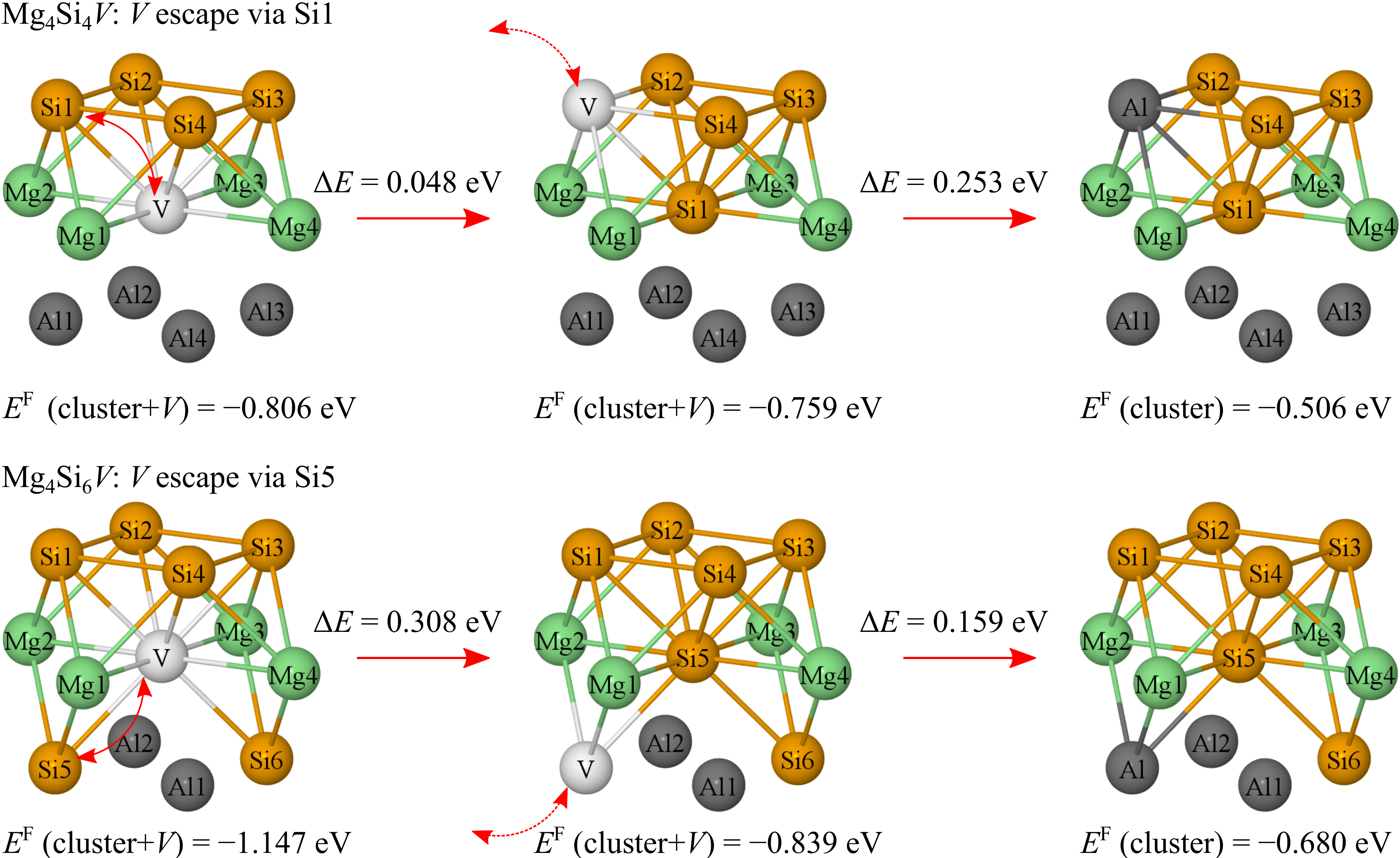}
\end{minipage}

\begin{minipage}[b]{.7\linewidth}
	\begin{table}[H]
		\centering
		\resizebox{5in}{!}{\begin{tabular}{l|ccccccc}
				
				\hline%
				\multirow{6}{*}{
				    \includegraphics[width=0.3\textwidth]{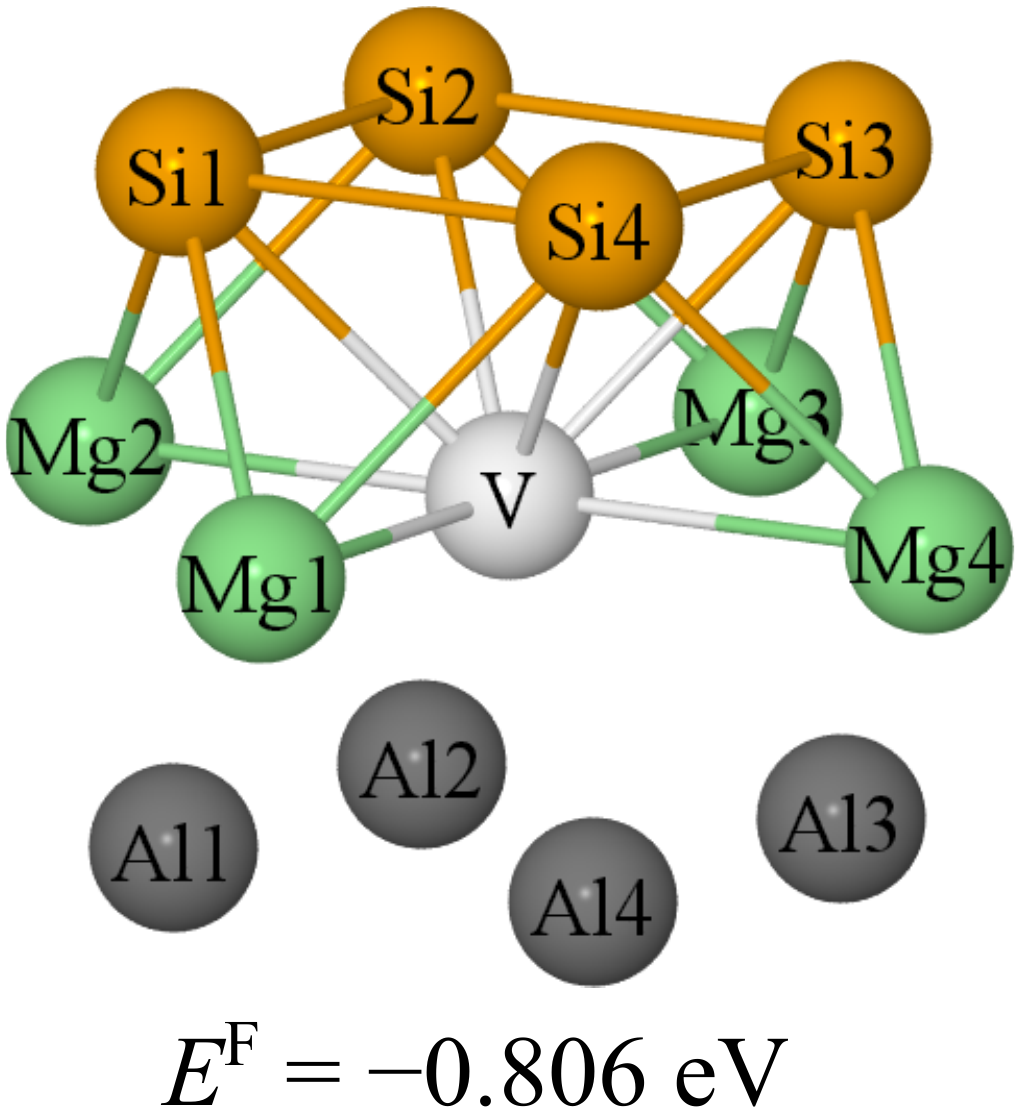}
                }   & \multicolumn{6}{c}{Energies [eV]}  &                   \\
				\cmidrule(r){2-7}
				&V path & $\text{Initial} $ & $\text{Intermediate} $  & $\text{Final} $ & Surface trap & Deep trap &\\
				\cmidrule(r){2-7}
				\cmidrule(r){2-7}
				& Si1 & $-0.806$ & $-0.759$ & $-0.506$  & $-0.253$ & $-0.301$ & \\
				& Mg1 & $-0.806$ & $-0.613$ & $-0.438$  & $-0.175$ & $-0.368$ & \\
				& Al1 & $-0.806$ & $-0.328$ & $-0.345$  & $-0.017$ & $-0.461$ & \\
				& $ $ & $ $ & $ $ & $ $  & $ $ & $ $ & \\							
				& $ $ & $ $ & $ $ & $ $  & $ $ & $ $ & \\		
				& $ $ & $ $ & $ $ & $ $  & $ $ & $ $ & \\										
								& $ $ & $ $ & $ $ & $ $  & $ $ & $ $ & \\		
				\bottomrule
				\hline%
				\multirow{7}{*}{
                    \includegraphics[width=0.30\textwidth]{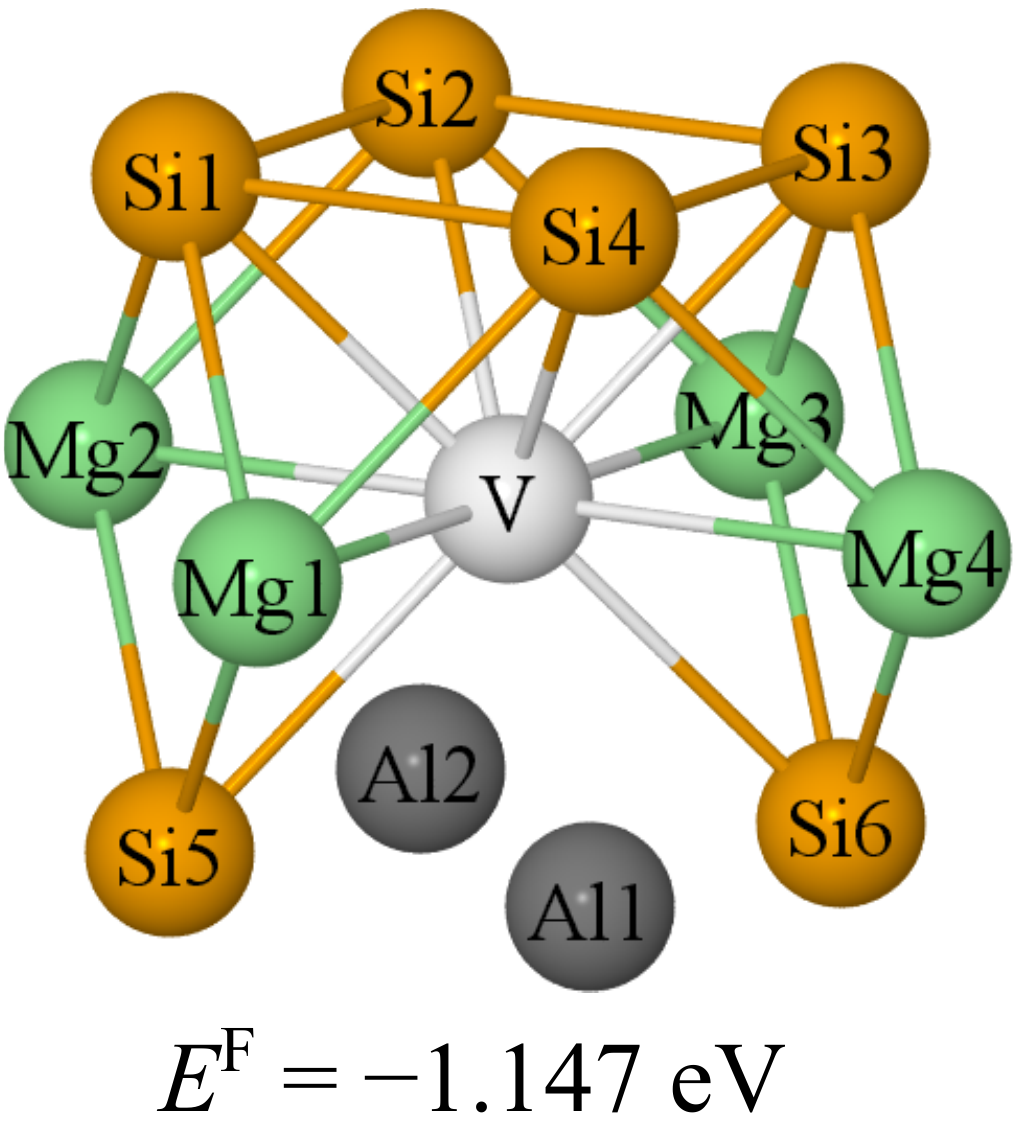}
                }   & \multicolumn{6}{c}{$ $}  &                   \\
				& Si1 & $-1.147$ & $-0.893$ & $-0.618$  & $-0.276$ & $-0.530$ & \\
				& Si2 & $-1.147$ & $-0.901$ & $-0.643$  & $-0.258$ & $-0.504$ & \\
				& Si5 & $-1.147$ & $-0.839$ & $-0.680$  & $-0.159$ & $-0.468$ & \\
				& Mg1 & $-1.147$ & $-0.832$ & $-0.582$  & $-0.251$ & $-0.565$ & \\
				& Al1 & $-1.147$ & $-0.687$ & $-0.522$  & $-0.165$ & $-0.625$ & \\
				& $ $ & $ $ & $ $ & $ $  & $ $ & $ $ &\\
								& $ $ & $ $ & $ $ & $ $  & $ $ & $ $ & \\		
												& $ $ & $ $ & $ $ & $ $  & $ $ & $ $ & \\	
																& $ $ & $ $ & $ $ & $ $  & $ $ & $ $ & \\			
				\bottomrule

		\end{tabular}}    
	\end{table}
\end{minipage}
	\caption{Top: atomic configurations for one path of vacancy escape from a cluster, for the 8-atom solute $\text{Mg}_{4}\text{Si}_{4}V$ cluster and the 10-atom solute $\text{Mg}_{4}\text{Si}_{6}V$ cluster, showing NNP energies of each configuration and energy changes when vacancy escapes via each of the neighbors. Bottom: Energy changes for V escape via all possible paths in the same 8- and 10-atom clusters.  All Si sites, Mg sites and Al sites are equivalent by symmetry in the 8-atom cluster.  Similarly, all Mg sites are equivalent in the 10-atom cluster and other site equivalencies are Si1 = Si3, Si2 = S4, Si5 = Si6, and Al1 = Al2.  }
	\label{fig10}
\end{figure}

We now examine the vacancy trapping energies for several typical clusters because total cluster energies with and without vacancies are not directly equivalent to the vacancy trapping energies due to changes in cluster configurations.  Assessment of vacancy trapping involves starting with a cluster containing a vacancy and then following possible low-energy paths by which the vacancy can escape from the cluster, which involves changes in local solute atom configurations.  The lowest-energy escape path determines the relevant trapping energy.  Figure \ref{fig10} shows vacancy traps and one possible escape path for typical 8- and a typical 10-atom solute clusters, respectively.  In both clusters, the vacancy initially has a maximum number of solute atom neighbors, residing on the surface of the 8-atom solute cluster (there is no interior) and residing in the interior of the 10-atom solute cluster.  The first step in the vacancy escape path examined here occurs by exchange with a neighboring atom, with Si shown in both cases.  After the exchange, the vacancy resides on the surface of the cluster in both cases (fewer solute atom neighbors).  In the 8-atom solute cluster, the energy change is small (0.048 eV) but in the 10-atom solute cluster case the energy change is large (0.308 eV). The second step in the escape is the exchange of the vacancy with an adjacent Al atom, and then subsequent vacancy-Al exchanges until the vacancy is far from the cluster (not shown).  The total energy change in this collective second step is moderate (0.253 eV and 0.159 eV, respectively) for both 8- and 10-atom solute clusters, but shows that the vacancy on the surface is also trapped.  The sum of the two energy changes is the apparent vacancy trapping energy $\textit{for escape via the chosen path}$, 0.301 eV for the 8-atom solute cluster and 0.467 eV for the 10-atom solute cluster.  This energy is also the difference in formation energy between the initial (vacancy in the cluster) and final (vacancy in the Al matrix) configurations. 

The table in Figure \ref{fig10} shows similar results for the energy changes for three possible paths of escape for the 8-atom solute cluster and 5 possible paths for the 10-atom solute cluster.  For each possible path, the ''surface trap" energy is the difference between the final and intermediate energies while the ''deep trap" energy is the difference between the final and initial energies. For the 8-atom solute cluster, the energy changes for exchanging the vacancy and either a Si or Mg solute atom are fairly small while exchange with an Al is costly and hence unfavorable. The vacancy will tend to migrate around to different Si/Mg solute atom positions in the cluster, with the original center position being most favorable.  The subsequent vacancy escape into the Al matrix, as indicated by the ''surface trap" energies, has barriers of 0.253 and 0.175, while the barrier for escape from the ``deep trap" energy is larger, (0.301 eV to 0.368 eV).  The vacancy will thus generally escape via the lowest energy path, which occurs via exchange with a Si atom, with a trapping energy of $\Delta E_t \approx 0.3$ eV.  At natural aging (room temperature), a typical time for escape of the vacancy is thus $t_{esc}=\nu^{-1} exp((\Delta E_t+\Delta E_{mig})/kT)\sim 50$ s.  Vacancy trapping for small clusters and surface sites will contribute to slowing natural aging, but is not responsible for the long trapping times observed in the kMC simulations containing a single vacancy.

For the 10-atom cluster, there are more pathways but similar trends. Escaping from the center of the cluster to the surface requires energies of $\approx 0.25$ eV for all the paths involving exchange with Si or Mg, so that the vacancy is fairly strongly trapped in the center position.  The subsequent escape into the bulk lattice (the ``surface trap" energy) is then comparable to that of the 8-atom cluster since the local environments of a surface vacancy are similar in the two cases.  The ''deep trap" energy is then notably larger for all of these paths, as compared to those for the 8-atom solute cluster, with the lowest energy path having a trapping energy of 0.468 eV.  A typical escape time for vacancy trapping in the center of such a larger solute atom cluster is $t_{esc} = \nu^{-1} exp((\Delta E_t+\Delta E_{mig})/kT)\sim 3.6 \times 10^4$ s.  This is significant on experimental time scales and consistent with the trapping times of $10^4$ s seen in the direct kMC (see Figure \ref{fig:enemg20}).

Finally, the study of compact clusters in Figure ~\ref{fig:fig4} showed the spontaneous formation of vacancy-Mg interstitial pairs, i.e. vacancies are created.  Although vacancies are an intrinsic part of the $\beta''$ structure (see Figures \ref{fig:fig3}a,c), it remains possible that the vacancies associated with the interstitial pair could escape into the Al matrix and then $\emph{enhance}$ solute atom diffusion and cluster/precipitate evolution.  We have examined this issue for the Mg$_5$Si$_9$ cluster.  The DFT-computed change in energy upon moving a Si from the Al matrix into the position of the vacancy, with the vacancy moved into the Al matrix, is 0.241 eV and so is energetically unfavorable.  These vacancies will thus prefer to remain with the clusters and, upon further cluster growth, will stabilize the emerging $\beta''$ structures and remain trapped in the interior of these structures. 

The full effects of trapping on delaying natural aging require a more complete analysis along the lines shown in ref. \cite{Francis2016}.  Since vacancies are constantly being trapped and released, the retardation of aging is associated with the average fraction of untrapped vacancies in the alloy that are available to transport solute atoms even though (many) other vacancies are trapped.  Previous work on trapping of vacancies by individual Sn atoms in Al-6xxx alloys \cite{Pogatscher2014DiffusionAlloys, Francis2016} showed that a Sn-vacancy trapping energy of 0.3 eV could retard natural aging by ~2 weeks.  Hence retardation of natural aging due to vacancy trapping on the surface of natural-aged clusters of sizes 8-10 atoms (examined explicitly here) can be expected.  Trapping in the interior of larger clusters can lead to an even longer delay, or even cessation, of natural aging.  A full calculation requires assessing the probability of finding the vacancy in the (one or a few) deep traps relative to the more numerous surface traps with slightly lower trapping energies, and this is beyond the scope of the present work.

 Qualitatively and quantitatively, our kMC studies and associated analysis clearly reveal the formation of ''vacancy prisons" \cite{Pogatscher2011} that arise for natural-aging clusters of sizes 8-15 atoms.  These vacancy prisons have been postulated to be the key to interpreting experimental results on natural aging kinetics.  As shown in  Figure \ref{fig:myfig}, the evolution of natural aging clusters increases up to aging times of $\approx 10^6$ s ($\approx$ 1 day) with the most frequent size range of 10-14 atoms, and then there are limited changes in cluster densities and limited coarsening over another factor of 10 in time scale (2 weeks or more).  Our kMC shows the same significant slowing of the aging with clusters of sizes 10-15 solute atoms, and that those clusters are able to trap solute atoms with trap energies consistent with the slowing of the natural aging.  Recalling that our kMC time scales are estimated, and rely on both an attempt frequency and an estimated quenched vacancy concentration, we do not expect quantitative agreement on the time scales but most other features of the experiments are captured well.

\section{Summary}

Kinetic Monte Carlo simulations of the early stages of natural aging in a near commercial composition of Al-Mg-Si alloy, using a recently-developed neural network potential, have demonstrated many features deduced or speculated based on meso- or macroscale experiments.  The kMC shows the formation of compact stable solute atom clusters with Mg/Si ratios and sizes consistent with inferences drawn from tomography experiments.  The lack of further cluster growth is explained by trapping of vacancies once the solute atom clusters grow to sizes of 8-15 atoms, with trapping energy of $~$0.3 eV and, less frequently but more importantly, $~$0.5 eV.  

The geometries of the stable clusters observed during kMC mainly stem from a pre-$\beta''$ motif, but with Si substituting for the central Mg (Figure \ref{fig:betamotifs}). Further stacking of these lower-energy units leads to the spontaneous formation of vacancy-Mg interstitial pairs that are the basis for the creation of $\beta''$ (Figure \ref{fig:fig4}).  {Nonetheless, the presence of Si atom(s) in central Mg site(s) (e.g. non-$\beta''$ Mg$_{9}$Si$_9$ rather than pre-$\beta''$ Mg$_{9}$Si$_9$) leads to significant lowering of the energy (-0.632 eV for this case as computed by DFT).  These non-$\beta''$ clusters are thus more stable and must be partially dissolved (Si replaced by Mg) to form the pre-$\beta''$ clusters that can eventually grow into $\beta''$ precipitates.  This may be the ''dissolution" process suggested by experiments in Figure \ref{fig:myfig}b.}

Further studies remain valuable.  The kMC studies could be pursued to longer times in larger systems.  Equally importantly, the kMC could be used to examine artificial aging at higher temperatures ($T=443$ K) to observe and quantify cluster dissolution and further $\beta ''$ nucleation and growth.  And the kMC could then be used to examine promising heat-treatment schedules that might create $\beta''$ nuclei and minimize natural aging and its deleterious consequences.  The kMC algorithm itself could be improved by using actual vacancy migration barriers as a function of local composition, which requires further validation of the NNP for those barriers.  The kMC algorithms could be optimized further, such as by cataloguing previously-visited local configurations and migration barriers rather than global geometries.  All such studies will increase the computational cost, and so must be carefully designed.

All of these results are enabled by the use of a near-chemically-accurate neural network potential (NNP) for Al-Mg-Si.  We have provided numerous new comparisons of the NNP to DFT for  a range of clusters to demonstrate the good accuracy of the NNP for studying natural aging in this alloy.  The NNP, while not perfect, is suitable for applications beyond kMC to explore longer time scales and larger length scales.  The NNP is also useful for modeling other aspects of alloy performance, with near-DFT accuracy with much lower computational cost, as demonstrated very recently \cite{Yi2022}.  More broadly, the machine learning framework and NNP methodology specifically are not limited to Al-Mg-Si alloys.  The development of new NNPs for other Al alloys (e.g. Al-Cu \cite{Marchand2020}, Al-Cu-Mg and Al-Mg-Zn-Cu \cite{Stemper2021}) and other alloys of high technological importance is on-going, and will enable the computationally-efficient study of many different problems in the processing and performance of alloys. 

\section{Methods}

We use the NNP16 \cite{Jain2021} for Al-Mg-Si designed particularly for Al-6xxx alloys. The potential is based on the formulation introduced by Behler and Parrinello \cite{Behler2015,Behler2007}.  The NNP was trained on the energies and forces computed by $ab$  $initio$ Density Functional Theory for a wide range of structures relevant for the metallurgy of this alloy.  The NNP16 was examined in considerable detail for many different atomistic configurations, including a few clusters posited to be important in NA \cite{Giofre2017} and emerging from earlier limited kMC studies. The NNP16 energies showed excellent agreement with DFT for a variety of configurations relevant to precipitation and natural aging, such as $\beta''$ and its compositional variants,  $\beta'$, formation energies of Mg, Si and vacancy in solid solution, antisite energies in various precipitates, and bulk vacancy transition state energies.  A preliminary natural aging study with NNP16 showed clustering of solute atoms at timescales commensurate with experimental observations \cite{Jain2021}, and the predicted cluster formation energies were in good agreement with DFT.  Here, we will further validate NNP16 against DFT calculations of clusters with and without vacancies that emerge commonly in the kMC study.  The DFT validation details are identical to those used in the development of NNP16 \cite{Jain2021}. 

In this work, the NNP16 potential is used to compute the energies of atomistic configurations that form during quasi-on-lattice kinetic Monte Carlo (kMC) simulations at a realistic alloy composition (Mg1$\%$Si0.6$\%$). The kMC simulations were carried out within the framework of the i-Pi code~\cite{Kapil2019ipi}, with the following details.  The kMC simulation cell contains a single vacancy, 17 Mg atoms, and 10 Si atoms in a 1728 atoms simulation cell with periodic boundary conditions.  The solute atoms and vacancy are added randomly; for such a large cell and dilute concentrations, the use of special quasi random (SQS) structures is not necessary.  The true supersaturated vacancy concentration is much lower ($10^{-5}$ - $10^{-6}$) than the vacancy concentration in the simulations ($c_V$ = 1/1728) because it is not feasible to model larger systems over the necessary time scales.  This difference in vacancy concentration accelerates kinetic phenomena in proportion to the vacancy concentration, and so a correction must be made to the kMC time scales.  All atoms and the vacancy are topologically related to a reference fcc lattice.  However, the actual atomistic system is relaxed to lower energy for a fixed number of molecular statics steps after each kMC step so that atoms are not restricted to lie on the exact lattice sites.  This enables relaxation of the solute atom clusters and the development of mechanical fields during the evolution.  These simulations are thus termed quasi-on-lattice.

The evolution of the atomic system is achieved by diffusion of the vacancy to one of its near-neighbor sites at each kMC step. For a given atomistic configuration with the vacancy at some initial fcc lattice site and system energy $E_i$, the energy changes $E_f-E_i$ upon moving the vacancy to each of the twelve possible final near-neighbor sites $f = 1,2,..12$ are first calculated. Retention of previously-computed atomistic configurations in a growing cache reduces the computational cost.  The migration barrier for each possible vacancy jump, $\Delta E^{M}_{i-f}$, is then approximated as
\begin{equation}
	\Delta E^{M}_{i-f}= \Delta E^{M}_{Al} + (E_{f} - E_{i})/2,
\end{equation}
where $\Delta E^{M}_{Al}=0.58$ eV is the bulk vacancy migration barrier in Al.  It is not currently feasible to compute the actual local barriers for every possible jump in every possible local atomic configuration, and so this approximation is necessary.  Furthermore, the controlling migration barriers for vacancy exchange with Si (0.52 eV) and Mg (0.68 eV)  in the Al lattice are not significantly different than that for pure Al, making this approximation reasonable.  Using these migration barriers, a catalog of rates $w_{i-f} = \nu\exp^{-\Delta E^{M}_{i-f}/k_{\text{B}}T}$ is created where $\nu= 1.66 \times10^{13}$ $\text{s}^{-1}$ \cite{Mantina2008} is an attempt frequency for all jumps.  The probability of jumping from $i$ to any $f$ is then $w_{i-f}/ \sum_{f=1}^{f=N} w_{i-f}$.  In practice, the transition selected at each step of the kMC is randomly sampled by building a cumulative function $ R_k = \{\sum_{f=1}^{f=k} w_{i-f}\} \text{ for } k = 1,2,..N$, selecting a random number $R$ from a uniform distribution in $(0,1]$, and selecting the event $k$ satisfying $R_{k-1} < R \leq R_{k}$. The event is the exchange of the vacancy with the position of the solute atom in the chosen $k$th configuration. The time step for the kMC move is then $t= -\log{R}/\sum_{f}w_{i-f}$.  A physical time for the jump is then obtained by scaling the kMC time by a factor of $10^6/1728$ to account for the difference between the real supersaturated vacancy concentration of $\approx 10^{-6}$ in the alloy and the vacancy concentration of $1/1728$ in the simulation.  This leads to a better contact between the kMC and timescales pertinent in real alloys.  

At every 1000th time step of the kMC simulation, we extract the total energy, atomistic configuration, and total root mean square vacancy displacement relative to the original vacancy position (and accounting for the system periodicity).  We further examine solute atom clusters defined as a connected set of solute atoms, i.e. solute atoms that share one or more near-neighbor with other solute atoms in the same cluster.  A minimum cluster size of 6 solute atoms was selected because smaller clusters were found to be very prone to frequent cycles of formation and dissolution.  The energies of selected solute atom clusters were recomputed using both DFT and NNP16 by extracting the cluster geometries from the 1728-atom cell, embedding them in periodic 108-atom FCC $3\times3\times3$ cubic supercells with all other atoms being Al, and minimizing the total energy. The shape and volume of the FCC supercell were kept fixed at the bulk Al FCC values. The cluster formation energy was then calculated as the cluster energy relative to the solid solution energy as
\begin{equation}
	E^{\text{F}}(\text{Mg}_{x}\text{Si}_{y})_\text{Al} = E(\text{Mg}_{x}\text{Si}_{y}\text{Al}_{N-x-y})- (N-x-y)E^{\text{F}}_{\text{Al}}-xE^{\text{F}}_{\text{Mg}_{\text{Al}}}-yE^{\text{F}}_{\text{Si}_{\text{Al}}}
\end{equation}  
where
\begin{equation}
	E^{\text{F}}_{\text{Si}_{\text{Al}}} = E(\text{Si}\text{Al}_{N-1}) -(N-1)E^{\text{F}}_{\text{Al}} 
\end{equation}  
\begin{equation}
	E^{\text{F}}_{\text{Mg}_{\text{Al}}} = E(\text{Mg}\text{Al}_{N-1}) -(N-1)E^{\text{F}}_{\text{Al}} 
\end{equation}  
with $N = 108$ and  $E(\text{Mg}_{x}\text{Si}_{y}\text{Al}_{N-x-y})$ the total energy of the 108-atom FCC supercell containing $x$ Mg and $y$ Si atoms.  Negative values of the cluster formation energy thus indicate increasing stability relative to the solid solution state.  Below we will make frequent comparisons of NNP and DFT formation energies to further support quantitative application of the NNP for natural aging. 

\section{Acknowledgments}
This research was supported by the NCCR MARVEL, a National Centre of Competence in Research, funded by the Swiss National Science Foundation (grant numbers 182892 and 205602).  The NNP16 potential was trained using N2P2~\cite{N2P2} and tested using LAMMPS~\cite{plim95jcp}. The kMC simulations were carried out within the framework of the i-Pi code~\cite{Kapil2019ipi}.

\section{Funding}
This research was supported by the NCCR MARVEL, a National Centre of Competence in Research, funded by the Swiss National Science Foundation (grant numbers 182892 and 205602).  

\section{Ethics Declarations}
The authors declare no conflicts of interest.

\section{Data Availability}
Data sets generated during the current study are available from the corresponding author on reasonable request.

\section{Author Contributions}
All authors contributed to the preparation of this manuscript. The simulations and data analysis were performed by Abhinav C. P. Jain. All authors read and approved the final manuscript.

\newpage
\appendix
\section{Appendix: Additional kMC trajectories}

\begin{figure}[H]
	\centering
	\includegraphics[width=6in]{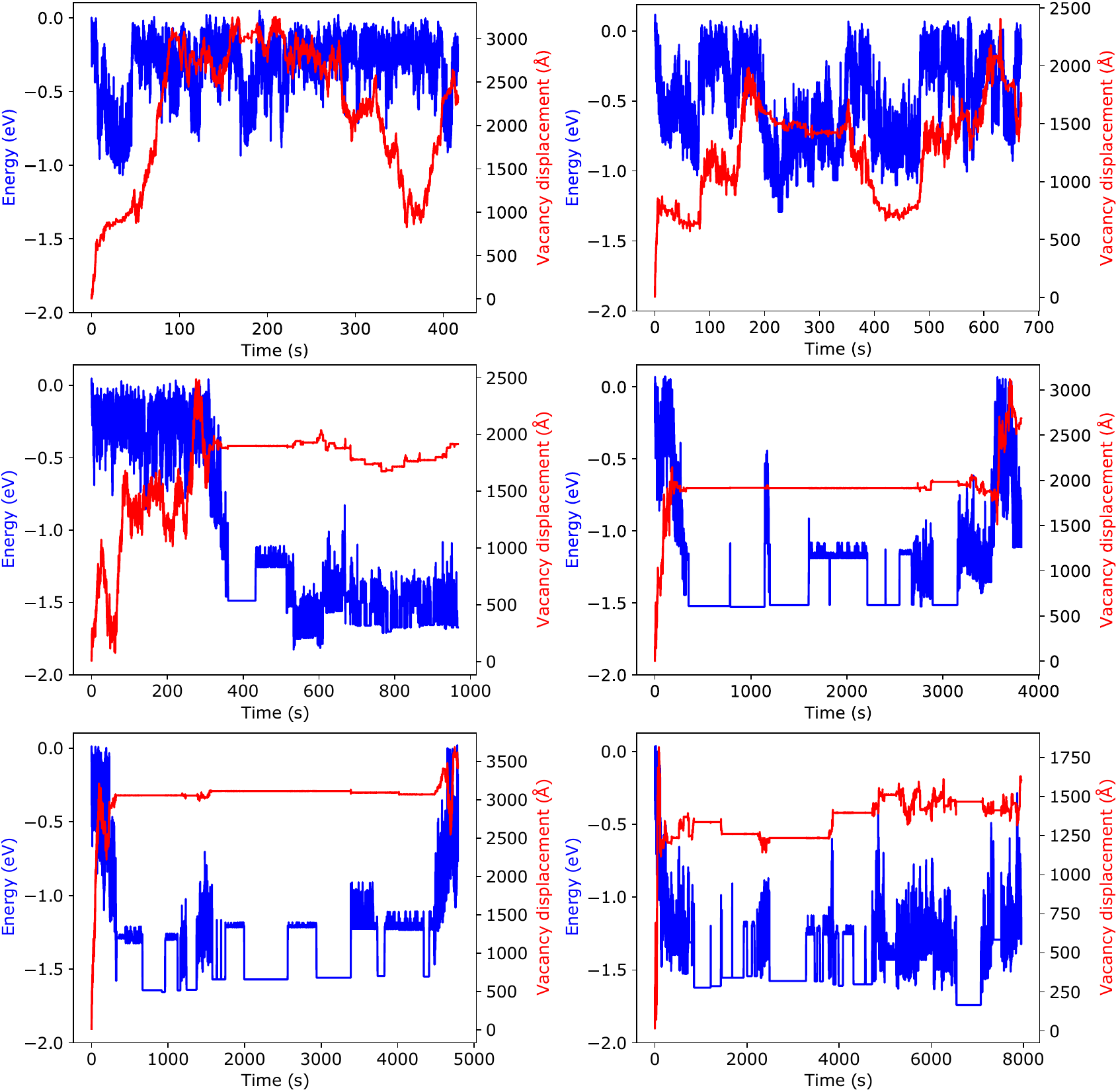}
	\caption{ kMC Trajectories which are still in early stages of simulation. 
	}
	\label{fig:traj}
\end{figure}

Figure \ref{fig:traj} shows trajectories of kMC simulations where clustering is still in the early stages and so nucleation-dissolution cycles can be observed. Some of the trajectories start showing cluster formation as evident from the drop in energies down to 1.5 eV. However, these clusters are not yet forming strong traps and hence the vacancy is able to more-readily escape. The total time accumulated by these simulations is then $1-2$ orders of magnitude smaller than the strongly-trapped trajectories illustrated in Figure \ref{fig:enemg20}. 

\end{document}